\documentclass[sigconf]{acmart} 

\AtBeginDocument{%
	\providecommand\BibTeX{{%
			\normalfont B\kern-0.5em{\scshape i\kern-0.25em b}\kern-0.8em\TeX}}}

\usepackage{subfigure}

\copyrightyear{2022}
\acmYear{2022}
\setcopyright{rightsretained}
\acmConference[CHI '22]{CHI Conference on Human Factors in Computing Systems}{April 29-May 5, 2022}{New Orleans, LA, USA}
\acmBooktitle{CHI Conference on Human Factors in Computing Systems (CHI '22), April 29-May 5, 2022, New Orleans, LA, USA}
\acmDOI{10.1145/3491102.3517738}
\acmISBN{978-1-4503-9157-3/22/04}

\PassOptionsToPackage{table,xcdraw}{xcolor}
\definecolor{gray}{rgb}{0.9, 0.9, 0.9}
\definecolor{lightgray}{rgb}{0.95, 0.95, 0.95}

\definecolor{redh}{rgb}{.7,0,0}

\begin{document}
	

\title{A Performance Evaluation of Nomon: A Flexible Interface for Noisy Single-Switch Users}

\author{Nicholas Bonaker}
\affiliation{
\institution{Massachusetts Institute of Technology}
\city{Cambridge}
\state{MA}
\country{USA}}
\email{nbonaker@mit.edu}

\author{Emli-Mari Nel}
\affiliation{\institution{University of the Witwatersrand}
\city{Johannesburg}
\state{Gauteng}
\country{South Africa}}
\email{emlimari.nel@gmail.com}

\author{Keith Vertanen}
\affiliation{\institution{Michigan Technological University}
\city{Houghton}
\state{MI}
\country{USA}}
\email{vertanen@mtu.edu}

\author{Tamara Broderick}
\affiliation{\institution{Massachusetts Institute of Technology}
\city{Cambridge}
\state{MA}
\country{USA}}
\email{tbroderick@mit.edu}

\renewcommand{\shortauthors}{Bonaker, Nel, Vertanen, and Broderick}
	
	\begin{abstract}
		Some individuals with motor impairments communicate using a single switch --- such as a button click, air puff, or blink. Row-column scanning provides a method for choosing items arranged in a grid using a single switch. An alternative, Nomon, allows potential selections to be arranged arbitrarily rather than requiring a grid (as desired for gaming, drawing, etc.) --- and provides an alternative probabilistic selection method. While past results suggest that Nomon may be faster and easier to use than row-column scanning, no work has yet quantified performance of the two methods over longer time periods or in tasks beyond writing. In this paper, we also develop and validate a webcam-based switch that allows a user without a motor impairment to approximate the response times of a motor-impaired single switch user; although the approximation is not a replacement for testing with single-switch users, it allows us to better initialize, calibrate, and evaluate our method. Over 10 sessions with the webcam switch, we found users typed faster and more easily with Nomon than with row-column scanning. The benefits of Nomon were even more pronounced in a picture-selection task. Evaluation and feedback from a motor-impaired switch user further supports the promise of Nomon.
		
	\end{abstract}

	
	\begin{teaserfigure}
\begin{centering}
    \includegraphics[height=72mm]{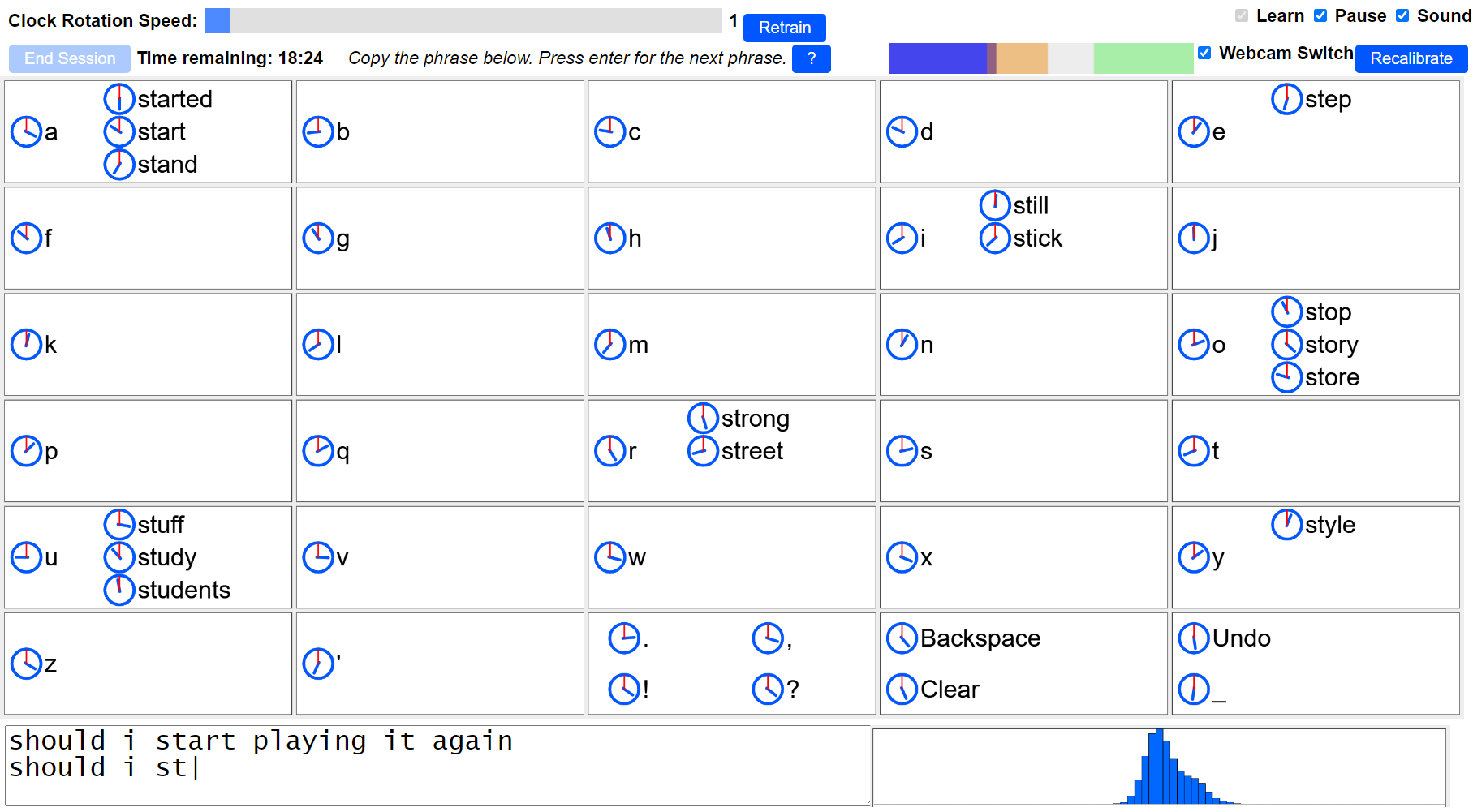} 
    \caption{Nomon being used to enter the phrase ``should i start playing it again''. The user has typed ``should i st'' thus far. To write the letter ``a'', the user clicks a switch when the clock near ``a'' is at noon. Target selection may require multiple clicks depending on the user's switch precision (visualized by the histogram). The user can select word completions to speed writing.}
    \Description{A screenshot of the Nomon Keyboard web application. The interface is divided into a five by six grid of boxes containing the letters a through z, punctuation marks (including apostrophe, period, comma, exclamation point, and question mark), a space character, and corrective options (including undo, backspace, and clear). In addition, the letter boxes contain up to three word predictions starting with the corresponding letter. Each option has a clock indicator immediately to its left.  }
    \label{fig_teaser}
\end{centering}
\end{teaserfigure}

	\maketitle

	\section{Introduction} \label{section:Introduction}

Individuals with severe motor impairments, such as cerebral palsy or locked-in syndrome, often communicate via augmentative and alternative communication (AAC) devices with single switch input \citep{leung2014autonomic,muller2013single,gibbons2010functional}. Users control the activation time of the switch by, e.g., pressing a button, releasing a puff of air, or blinking \citep{angelo2000factors,grauman2003communication,grafton2010unlocking}. Most commonly, these switch activations (henceforth "clicks") are used as input to a scanning interface \citep{simpson2011modeling,simpson2006evaluation}. The graphical user interface highlights different options in turn; the interface chooses whichever option is highlighted when the switch is activated. But highlighting every option in sequence can be inefficient for even a moderate number of options. While a popular variant called row-column scanning is more efficient, it requires that options be arranged in a grid. Computer users often need to choose among options not arranged in a grid; e.g.\ in drawing, gaming, and web browsing.\footnote{Currently switch users are primarily limited to games and websites that have certain, constrained switch-friendly formats; see, e.g., \url{https://www.bltt.org/switch/activities.htm} and \url{ https://everydayspeech.com/adventures-switch-accessible-websites/}.}

Nomon \citep{nomonjournal,nomonthesis} offers a more flexible user experience. Nomon places an indicator next to each selection option, and uses a probabilistic selection mechanism to avoid inefficiently visiting each option in turn. Previous research suggests users can type more quickly and easily using Nomon than using row-column scanning \citep{nomonjournal,nomonthesis}. 
However, there were four main deficiencies in past work: 
\begin{enumerate}
\renewcommand{\theenumi}{\Alph{enumi}}
\item Previous research tested each user for less than an hour total using Nomon. Performance may differ with more experience. 
\item In past research, non--motor-impaired users triggered Nomon using a joystick button. In this paper, we refer to individuals who do not regularly use AAC switches as ``non--switch users.'' Motor-impaired individuals often exhibit different single-switch reaction times relative to non-switch users.\footnote{See the first row of Figure \ref{figure:Reaction Times Plot} (Section \ref{section:approximating_motor_impairments}) below, showing data kindly provided by Dr.\ Heidi Koester and collected in \citep{koesterpsycometric,koesterscanwizzard}.}
\item Previous research had users enter phrases consisting primarily of common words. Such phrases are easier to write in Nomon due to its use of a language model and the interface's word completions. But a text entry method also needs to support the input of uncommon words (e.g., proper names). 
\item Previous research tested only text entry. But Nomon promises to enable efficient input for tasks beyond text entry.
\end{enumerate}

Not only do we address these concerns in the present study, but we also improved the Nomon program to increase accessibility before running our study. To improve the Nomon program, we first consulted with AAC experts\footnote{The AAC experts we consulted include staff at charities SpecialEffect \url{https://www.specialeffect.org.uk/} and the Ace Centre \url{https://acecentre.org.uk/}.} and a single-switch user. Based on their feedback, we designed a more accessible interface for Nomon. Moreover, our AAC consultants flagged that the original Nomon initialization was likely to require impractical or costly manual interactions. So we developed a more suitable initialization process. Further, we adapted simulation methods used to optimize internal parameters in scanning systems \cite{koester1994modeling, simpson2011modeling, koesterrcsmodel} to design a more efficient keyboard layout for Nomon.

After these updates, we addressed concern (A) by performing a longer user study; we collected data on each study participant's use of both row-column scanning and Nomon across 10 sessions. 

We addressed (B) in two parts. First, we tested the performance of a single-switch user\footnote{The user we consulted about interface design and the user whose performance we tested were two different single-switch users.} with the Nomon interface.
Second, recognizing the especially valuable time of motor-impaired users, we focused our larger-scale testing on non--switch users. But crucially we developed and validated the use of a webcam-based switch to allow non--switch users to better approximate reaction times of motor-impaired users. Although our approximation is not fully representative of all single-switch users, our results in the present paper employ click timings that are better aligned with the target population than the original Nomon study. This approximation allowed us to develop and test our new initialization method and the general effect of a noisier switch with non--switch users.  Our results were also useful to convince our collaborating charities that a study involving single-switch users would be worthwhile since a practical initialization procedure was deemed critical.

To address (C), we selected phrases so that a third contained a challenging word not in our language model's vocabulary. With this diverse sent of phrase prompts, we could not only measure user performance with relatively simple text, but also explore whether Nomon degrades gracefully in the face of  harder-to-predict text.

To address (D), we also compared performance of row-column scanning and Nomon in a task beyond text entry. There are many potential uses of Nomon such as gaming \citep{aced2015gnomon,lopez2017design,lopez2015playable,aced2016clocks,aced2015can}, drawing (\citep{nomonthesis}, Section 7.1), and general operating system control (\citep{nomonthesis}, Section 7.3, Section 2 of our supplement). But to facilitate comparison, we focus on a task where row-column scanning can still be applied: selection from a large set of pictures. We expect similar behavior when selecting among files on a desktop, selecting a computer application to launch, or selecting products at an online retailer. 

Our results demonstrate that, under these conditions, users find it faster and easier to enter text using Nomon than using row-column scanning. In the text-entry task, participants typed 15\% faster with Nomon and rated it easier to use. The benefits of Nomon are even more pronounced in the picture-selection task where participants selected targets 36\% faster. We make the following contributions: 
\begin{itemize}
    \item An updated and easily available Nomon interface, redesigned to increase accessibility through feedback from switch users and AAC specialists.
    \item A model of a Nomon user and a subsequent simulation study to optimize the design of the Nomon interface.
    \item A user study comparing non--switch users' performance with Nomon and row-column scanning in: (1) a text entry task with challenging out-of-vocabulary words and (2) a picture-selection task to simulate applications beyond text entry.
    \item A user trial of Nomon in a text-entry task with a motor-impaired switch user.
    \item A method for approximating motor-impaired reaction times with non--switch user inputs, and a validation of this method.
\end{itemize}

The rest of this paper is structured as follows. We survey approaches to single-switch text composition in Section \ref{Section:related_work}. We detail how our two interfaces operate, and justify our interface and study design choices, in Section \ref{section:Interface design}. We describe our user study and picture-selection task in Section \ref{section:User Study}. We describe our method for approximating reaction times of motor-impaired users in Section \ref{subsubsection:Apparatus and Software} and formally justify it in Section \ref{section:approximating_motor_impairments}.


	\section{Related work}
\label{Section:related_work}
\subsection{Input via scanning}

Individuals with motor impairments tend to write slowly using row-column scanning (RCS):
\citet{Koester2014enhancing} observe entry rates of 0.3--2.9 words per minute (wpm), and \citet{RoarkHuffman2015} observe 1.9~wpm. Researchers have investigated various approaches to speed text entry. Arranging letters cleverly in the grid can speed selection \cite{crochetiere1974computer,venkatagiri_efficient}. Carefully configuring the scanning interface can substantially impact performance \citep{Lesher1998, Angelo1992, Koester2014enhancing, koesterpsycometric, koesterscanwizzard,reactiontime}. Instead of scanning rows and columns sequentially, the interface can highlight subsets of cells in some way, e.g. via Huffman coding~\cite{baljko_indirect,roark_huffman,RoarkHuffman2015} or a language model~\cite{wandmacher_sibylle}.  Character or word predictions may speed input \cite{trnka_user,lesher1998techniques}, but not in all cases \cite{koester1994modeling,Koester1996,Koester2014enhancing}. 
Scanning can be used for applications other than text input, e.g.~navigation in virtual environments~\cite{folmer_3d} or playing games~\cite{yuan2011game}. However, despite these efforts, scanning requires choosing either (1) a fast scan rate that risks false selections or (2) precise target selection but with a slow scan rate. A further obstacle is that applications must be designed to fit the scanning paradigm, e.g.\ by placing options in a grid.

\subsection{Selecting a moving target within a time window}
Selecting a moving target within a particular time window is a task that arises in some situations, e.g., smartphone games.  For this task, researchers have studied error rates and models of timing performance \cite{lee_spatial_error,lee_temporal_error} as well as automated game playtesting \cite{lee_playtesting}. Controlling a timed activation in these cases (e.g., via a button or screen press) might be interpreted as an activation of a single switch. As it stands, though, the task remains distinct from either RCS or Nomon. In particular, in the moving target task, there is a continuous movement of some visual element, and the goal is to click when the element is in some spatial range, implying a particular time range for selection. In RCS, there is a fixed time window for making a selection, but there is no continuous visual element; instead the user is shown only the discrete highlighting of rows or columns. Nomon, by contrast, presents a continuous visual element (the rotating clock hand), but there is no fixed time window during which clicking realizes some goal. Rather, in Nomon, changing the timing of a user's click induces a continuous change in the observed likelihood value, and the likelihood shape itself is user-specific. This shape need not be Gaussian or even unimodal, and is learned by the Nomon method. It would be interesting to explore whether using a continuous visual prompt inspired by gameplay might aid, e.g., in the usability of RCS interfaces.

Additional work in this vein  models the neuromechanical process of a finger pressing a physical button in a non--switch user \citep{lee_impact_activation, lee_neuromechanics}. Single-switch users, though, often have specialized switches that may be activated by different body parts. For example, some switches detect puffs of air, electrical muscle activations, or blinking. It remains to be seen if modeling techniques in the same spirit might by usefully applied to these other forms of switch input.

\subsection{Input via a noisy switch}
In the Dasher interface, users write by navigating through a world of nested letter boxes \cite{ward_acm}. The size of each letter's box is based on language model that adapts as the user writes. Typically Dasher is operated via a pointing device such as a mouse or eye-tracker \cite{tuisku_dasher,rough_dasher}. Dasher can also be controlled via a single switch \cite{MacKayButtons2004, MacKayBall2006}.
Dasher applies Shannon's noisy-channel coding theorem \cite{shannon48} to facilitate efficient text entry. Dasher's  navigation interface allows selection of multiple letters or even entire words with a single click. This mechanism can reduce the physical effort and time required of users.
A pilot study showed a non--switch-using expert could write at 10\,wpm using only 0.4 clicks per character~\cite{MacKayBall2006}. To our knowledge, there have been no further user studies of one-button Dasher. \citet{MacKayButtons2004} note that the capacity of the channel is substantially reduced by a noisy switch with an erroneous activation a fraction $f$ of the time. Also Dasher requires options be arranged in a strict order (e.g., alphabetically) which can limit applications beyond text entry.

Unlike Dasher, Nomon does not require an ordering on selection options. Nomon also explicitly models, and nonparametrically learns, a distribution describing how a user clicks relative to a baseline time.
Dynamically adapting to the user's particular clicking style, represented by this click-time distribution, is a novel aspect of Nomon in the context of single-switch text entry methods. Theoretically, adaptation should result in less error correcting and faster text entry. A game for children with motor impairments \cite{aced2015gnomon,lopez2015playable,lopez2017design} demonstrates Nomon's applicability in real life. \citet{Nel2019} extended Nomon's noise model to develop a communication method for single-switch users who are also visually-impaired.

\citet{williamson2020efficient} presented a probabilistic user interface for binary input devices with high noise levels when reliability can be ensured only 65\%\textendash 90\% of time (e.g., non-invasive EEG).
Like Dasher, the interface works by progressively zooming in and draws heavily from information theory. The authors use Hornstein error correcting~\cite{AHN2015103} to increase noise tolerance. Their method builds up certainty for a sequence of user selections before making a decision on all the selections rather than deciding one target at-a-time.

	\section{Interface Design} \label{section:Interface design}

We use two interfaces in our study, Nomon and row-column scanning. Here we describe the interfaces in detail and justify our parameter choices in each case --- via both simulation studies and collaboration with AAC users and specialists. We also discuss how the COVID-19 pandemic affected our interface and user study.
    
\begin{figure*}[ht]
	\centering
	\includegraphics[width=135mm]{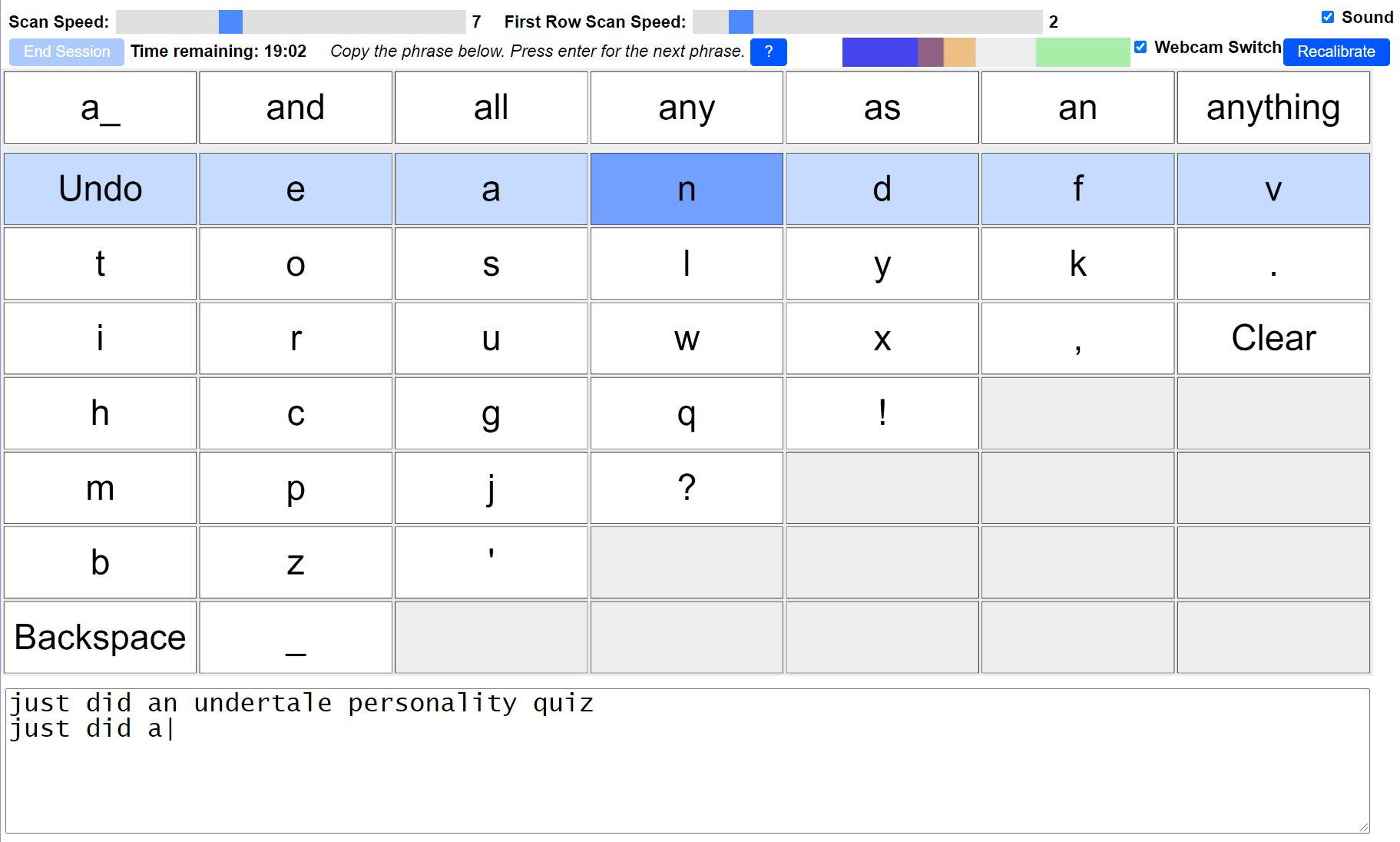}
	\caption{Our row-column scanning keyboard interface. The user is being prompted to write ``just did an undertale personality quiz'' and has written ``just did a'' so far. The interface progressively highlights rows until the user clicks their switch, and then progressively highlights columns within the selected row until the user clicks  once more. Users can also select word completions that are displayed at the top.}
	\Description{A screenshot of our row-column scanning keyboard interface. The interface is divided into a seven by eight grid of boxes containing the letters a through z, punctuation marks (including apostrophe, period, comma, exclamation point, and question mark), a space character, and corrective options (including undo, backspace, and clear), and 7 word completions on the top row. Items are arranged in an upside-down stair step fashion towards the top left of the grid. }
	\label{figure:Row Column Scanning Keyboard Interface}
\end{figure*}
    
\subsection{Row-Column Scanning}
\label{subsection:Row Column Aparatus}

\subsubsection{Background.} A row-column scanning (RCS) interface presents the user with a 2D grid of options. For a text entry task, these options are letters and word completions. The system scans through each row at a constant time interval called the scan delay. When a user clicks their switch, the interface selects the currently highlighted row and proceeds to scan through each column. The user clicks again when the column scan highlights their target. The second click makes a selection.

\subsubsection{Our implementation.} Figure \ref{figure:Row Column Scanning Keyboard Interface} shows our RCS implementation. While there are research and commercial RCS implementations \cite{grid_3,proloquo2go,novachat,coughdrop}, we implemented our own version since our goal was to compare the RCS and Nomon interfaces as directly and fairly as possible. Having our own implementation allows us to use the same word prediction engine in both interfaces and augment both interfaces with similar logging and experimental controls. As noted in \cite{Lesher1998,trnka_user}, word predictions can profoundly impact the entry rate and click load of switch users. As advised by the AAC consultants, we followed The Grid 3 (a popular commercial scanning software) design for our RCS interface. 

In both our RCS and Nomon text entry interfaces, the principal options were: character keys (the letters a--z); space; punctuation keys (comma, period, apostrophe, question mark, and exclamation point); and three correction keys --- undo (to revert the latest selection), backspace (to delete the current final character), and clear (to clear all text that currently appears).

In the event a user selects a row in error, we follow the recommendation of \cite{simpson2011modeling} and set the maximum number of column scans to two complete cycles. After this point, the interface reverts back to row scanning. This procedure stands in contrast to alternative options, such as requiring the user select an option to stop scanning the columns or to reverse the direction of scanning \cite{simpson2011modeling}. 

\subsubsection{Keyboard Layout}\label{subsec:rcs layout design}
Proper RCS configuration is critical for fast writing speeds \cite{Lesher1998, Angelo1992, Koester2014enhancing, koesterpsycometric, koesterscanwizzard, Koester1996, simpson2011modeling, venkatagiri_efficient}. Therefore, we ran a simulation study to determine optimal interface parameters in our RCS implementation; see the supplemental materials for full details. We then verified our results with the recommendations of the previous literature. Namely, we considered a maximum number of word completions to display at any one time $W_{\text{max}} \in \{1, 2,\dots ,18\}$. We considered whether to display word completions at the top or bottom of the interface. We also considered whether to sort character options alphabetically or by frequency (with more-common letters in English near the top left of the grid to reduce scan time to reach them). Optimizing text-entry rate in our simulations led us to choose to arrange letters in frequency order and to include seven word completions arranged by decreasing probability in the top row. The frequency arrangement coincides with the recommendation of \citep{venkatagiri_efficient, simpson2011modeling}. Our word completion arrangement matches the recommendation of \cite{koesterscanwizzard}.
The resulting grid was $8 \times 7$ in size with at most 42 options.

\begin{figure*}[tb]
    \centering
    \includegraphics[width=0.75\linewidth]{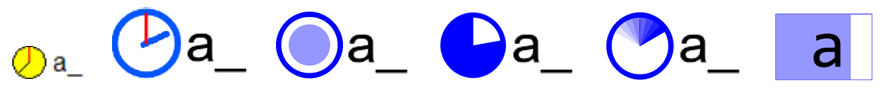}
    \caption{From left to right: (1) The clock used in the original Nomon interface \cite{nomonjournal}, (2) the clock design used in this study, (3) the filling ball clock, (4) the ``pac-man'' clock, (5) the radar clock, (6) the progress bar. Clock sizes are relative to the clock from the original interface on a \boldmath$2560 \times 1600$ pixel display.}
    \Description{An image showing the six indicator designs considered for Nomon. First, a small, yellow, thin-bordered clock with a red line at noon and a black minute-hand. All the following indicators are twice as large as the first. Second, a blue, thick-bordered clock with a red line at noon and a thick, blue minute-hand. Third, a thick-bordered, blue circle that fills with a light blue circle from the center to the edges. Fourth, a circle that fills in blue in a pac-man shape. Fifth, a blue, thick-bordered clock with a red line at noon and a blue minute-hand with radar trails. Last, a progress-bar-like box that fills blue horizontally.}
    \label{figure:clocks}
\end{figure*}

\subsubsection{User-adjustable parameters.}
As is common in RCS implementations \citep{koesterscanwizzard}, users could
control two timing parameters: the scan time and the extra delay. The scan time $s$ is how long the interface highlights an individual row or column. We set $s = 2e^{-j/14}$ seconds for $j \in \{0,1, \ldots, 20\}$. That is, smaller values of $j$ correspond to longer scanning delays with $s$ ranging between $[0.48, 2.00]$ seconds. The extra delay $d$ is added to the scan time for the first row and column. We set $d = 0.15(10-k)$ seconds for $k \in \{0,1, \ldots, 10\}$. Therefore, $d \in [0, 1.5]$ seconds. Smaller values of $k$ correspond to longer extra delays and $k=10$ corresponds to no extra delay. Participants started with the slowest settings of $j, k=0$ and were allowed to increase or decrease either $j$ or $k$ or both by 1 between phrases.

\subsection{Nomon}
\label{subsection:Nomon_Aparatus}

\subsubsection{Background.}
In Nomon, every option in the interface has a clock next to it (Figure \ref{fig_teaser}). Each clock has a unique phase, and the minute hands of all clocks rotate at a constant, shared speed. A user needs to look at only one clock to select its corresponding target. RCS and other methods are potentially more taxing in that they require a user to shift visual attention between different parts of the screen. The Nomon user is instructed to click when their target clock's hand passes the red ``noon'' line. After each click, the clock hands change phase. The phase change is chosen to separate the clock phases of the most probable next targets from one another. The user repeatedly clicks, each time aiming for when the minute hand passes noon, until their target is selected. The number of clicks required to select a target is dependent on the precision of the user and on how probable the target is. In a text entry application that makes use of a language model, an experienced user can select targets in around two clicks~\cite{nomonjournal}. A video demonstration of how typing with the Nomon interface works can be found in our supplemental materials.

\subsubsection{End-User and AAC Consultant Involvement in the Design Process}

Throughout the process of redesigning the Nomon interface, we consulted with two charities specializing in individuals with severe motor impairments: SpecialEffect and the Ace Centre. We received feedback from ten of the SpecialEffect staff members and one consultant from the Ace Centre. In addition, a single-switch user affiliated with SpecialEffect gave us feedback on usability and accessibility. All the feedback played a major role in our design choices, including: color options (e.g., to help prevent seizures or migraines), clock design, font choice, text contrast, the addition of a tutorial and calibration phase, and improved visual/audio selection feedback.

Our AAC-charity consultants noted that it can be cumbersome and error-prone to initialize parameters before using an AAC interface. In the case of Nomon, the click-time distribution estimate itself requires initialization. And we know that click-time distributions can vary considerably across users; see Section~\ref{subsubsection:Apparatus and Software}. So the fixed initialization of the click-time distribution estimate in the original Nomon would generally be misspecified for a new user. This discrepancy could necessitate impractical or costly manual intervention from carers or users. We therefore introduced a calibration phase to initialize the estimated click-time distribution of a user before they start using Nomon.  

At our consultants' suggestion, we also considered alternative indicators besides clocks; namely, from right to left in Figure~\ref{figure:clocks}, progress bars, clocks with radar trails, a ``pac-man'' filling clock, and filling circles. Based on the consultants' feedback, we settled on a larger clock design with thicker borders and higher contrast, and a larger, bolder font. These changes are consistent with modifications to Nomon to increase usability reported in \citep{aced2015gnomon}.

\subsubsection{Simulating a Nomon User.}
\label{subsubsection: simulating a nomon user}

\paragraph{Motivation.}
In the Nomon keyboard interface, two parameters control the presentation of word completions on the screen: $W_c$, the number of word completions in each character's box; and $W_{\text{max}}$, the total number of word completions allowed across all characters. In the original study of Nomon, $W_c$ was set to $3$ words per character as it was the maximum number that could fit on the screen, and $W_{\text{max}}$ was left uncapped \cite{nomonthesis}. 
Given prior success in optimizing parameters in scanning systems via simulation \cite{koester1994modeling, simpson2011modeling, koesterrcsmodel}, we investigated optimizing these two parameters in Nomon. We developed a model of a switch user that simulates text composition in the Nomon keyboard. We then applied this model to generate synthetic user selection data that could predict performance at various parameter configurations. 

We emphasize that Nomon's internal model (i.e., the model controlling when a selection is made, how the clock phases are set, etc.) is distinct from our user simulation. To emphasize the difference, note that Nomon can operate, with its own internal model, without employing a user simulation to optimize these two parameters; for instance, the parameters could instead be given default values. Conversely, we can (and do) employ a user simulation to optimize the parameters of RCS even though RCS does not include an internal model to control its operation like Nomon does.

\begin{figure}[tb]
	\centering
	\includegraphics[width=\linewidth]{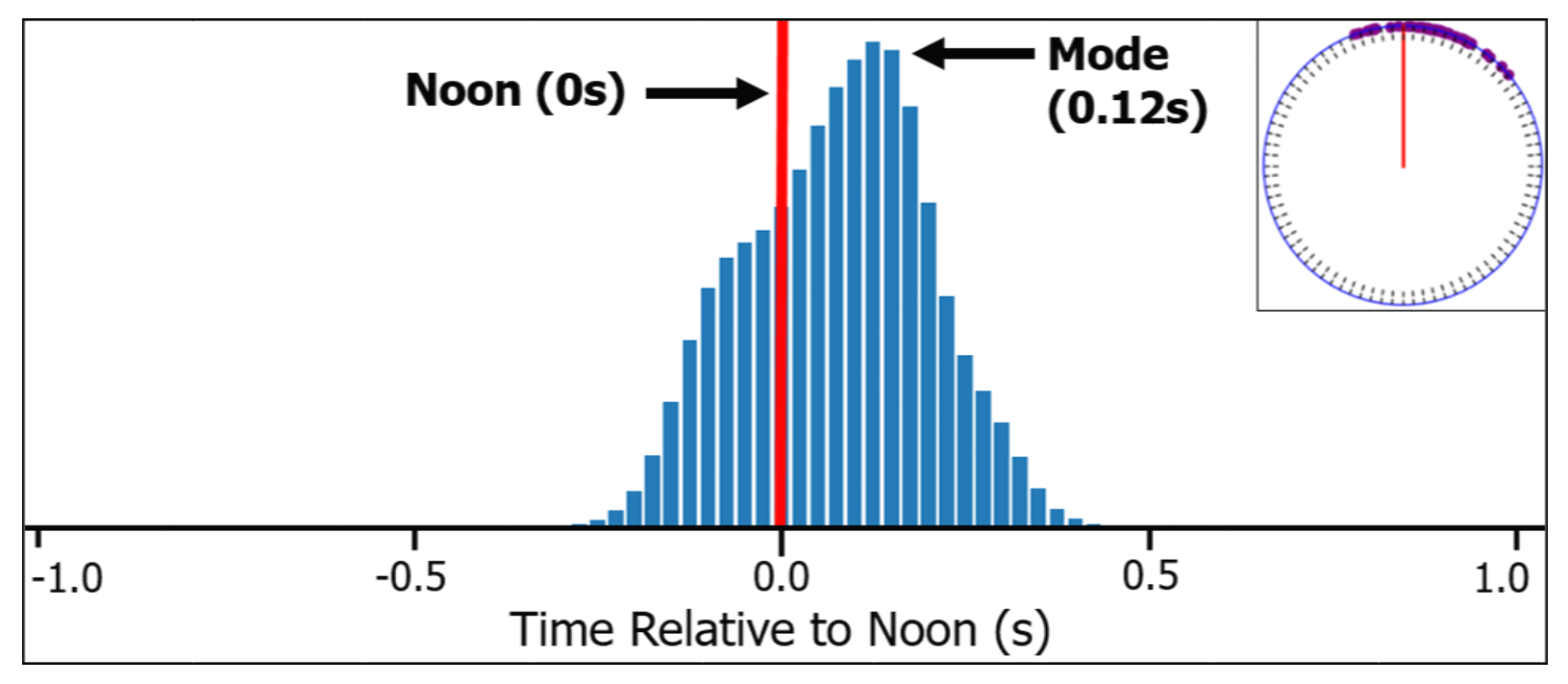}
	\caption{Visualization of a click-time distribution. A click-time distribution is a likelihood estimated by Nomon of when a user clicks relative to noon. The histogram above can be thought of as unraveled from the clock in the top right. The purple points around this clock are the samples (timings of clicks relative to noon) used to construct this likelihood estimate. The mode of this histogram occurs 0.12 seconds later than overlap of the moving hand with noon.}
	\label{figure:Click Dist}
	\Description{A blue histogram of click-timings relative to noon (zero seconds on the x-axis). There is a peak slightly after noon at 0.12 seconds. In the top right is a circular plot showing the locations of the hour hands for each data point used to generate the histogram. }
\end{figure}

\paragraph{Input Error Model: The Click-Time Distribution.}
A click-time distribution is a likelihood estimated by the Nomon interface of when a user clicks relative to noon. The histogram in Figure \ref{figure:Click Dist} visualizes this distribution for a particular Nomon user. Nomon estimates this distribution as part of its probabilistic selection mechanism \cite{nomonthesis}. After a series of clicks from the user, the posterior probability of each clock can be calculated through Bayes’ theorem. A clock is selected in Nomon if its posterior probability is sufficiently high \cite{nomonthesis}. As a consequence, if a user's click-time distribution is narrow, they will generally be able to select a clock with few clicks. If their distribution is wide, it will take more clicks to select an option. Precise users will be able to select clocks quickly, and imprecise users will require more clicks to to reduce the chance of an erroneous selection.

\paragraph{User Model.} Our user model interacts directly with a running instance of Nomon, receiving information on the available clocks and their hour-hand locations, and outputting switch activations. Given a phrase, the user model attempts to type the phrase by targeting and selecting character clocks and word completion clocks (if available) in the interface. It selects these target clocks by calculating the time until the target clock is at noon and simulating a switch event at that time.

However, real users rarely click exactly at noon. We added input noise to the user model by sampling from an experienced user's fixed click-time distribution. A click-time distribution sample was added as an offset to when the user model simulated a switch event. 

We then generated synthetic user data by running our user model on the corpus of phrases used for the text entry task in our user study (Section \ref{subsubsection:text entry proceedure}). We repeated these simulations while varying the two parameters that control the word prediction layout.

\paragraph{Keyboard Layout and Simulation Results}
Figure \ref{fig_teaser} shows the Nomon interface we used for our text entry study.
We divided the layout into a 6\,$\times$\,5 grid of principal options, with characters alphabetically arranged. Each character option could have a maximum of 3 word completions displayed next to it in the grid. Thus, there could be up to $3 \cdot 26$ word completions in total. 
\begin{figure*}[ht]
    \centering
    \includegraphics[width=70mm]{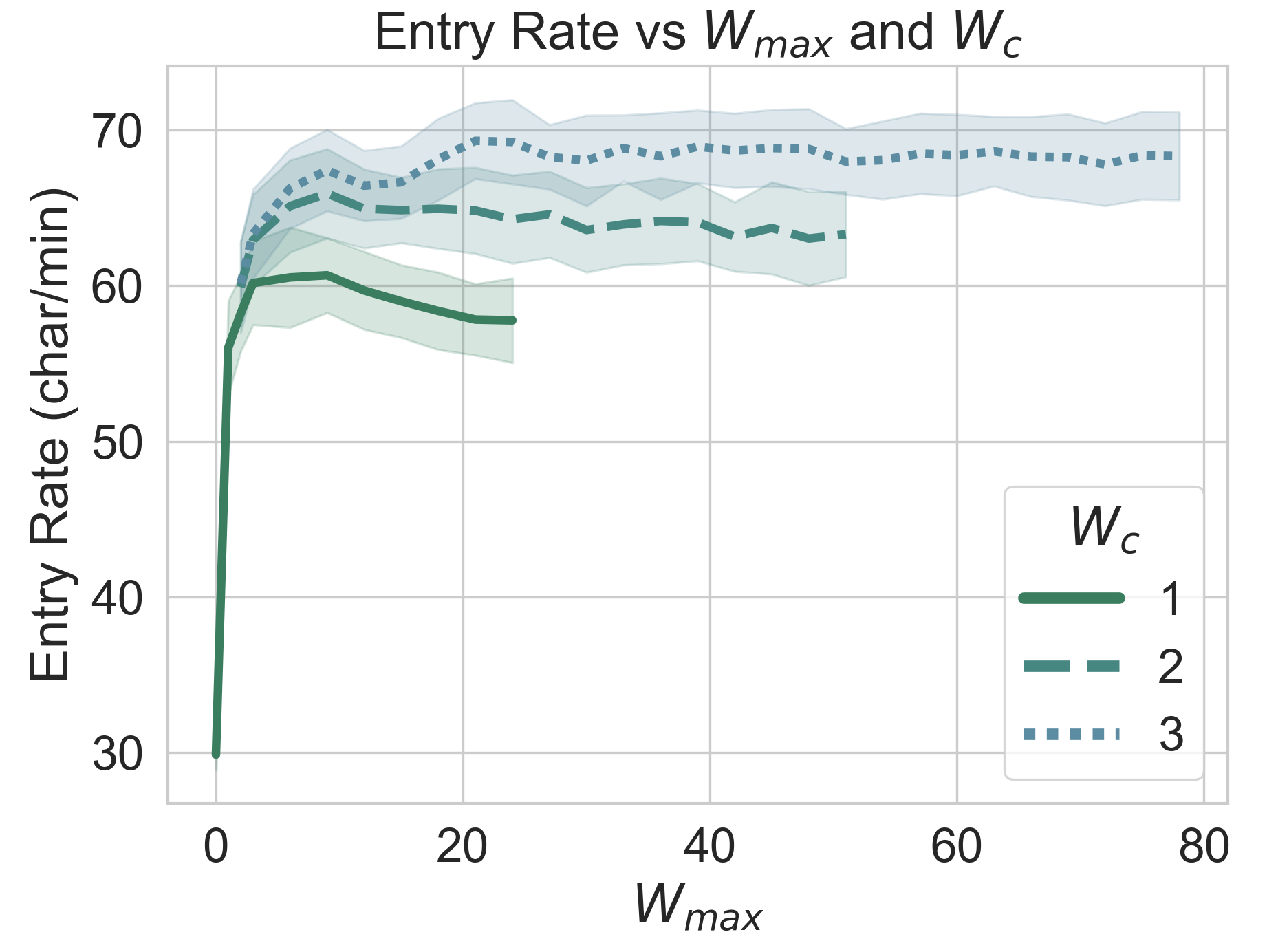}
    \includegraphics[width=70mm]{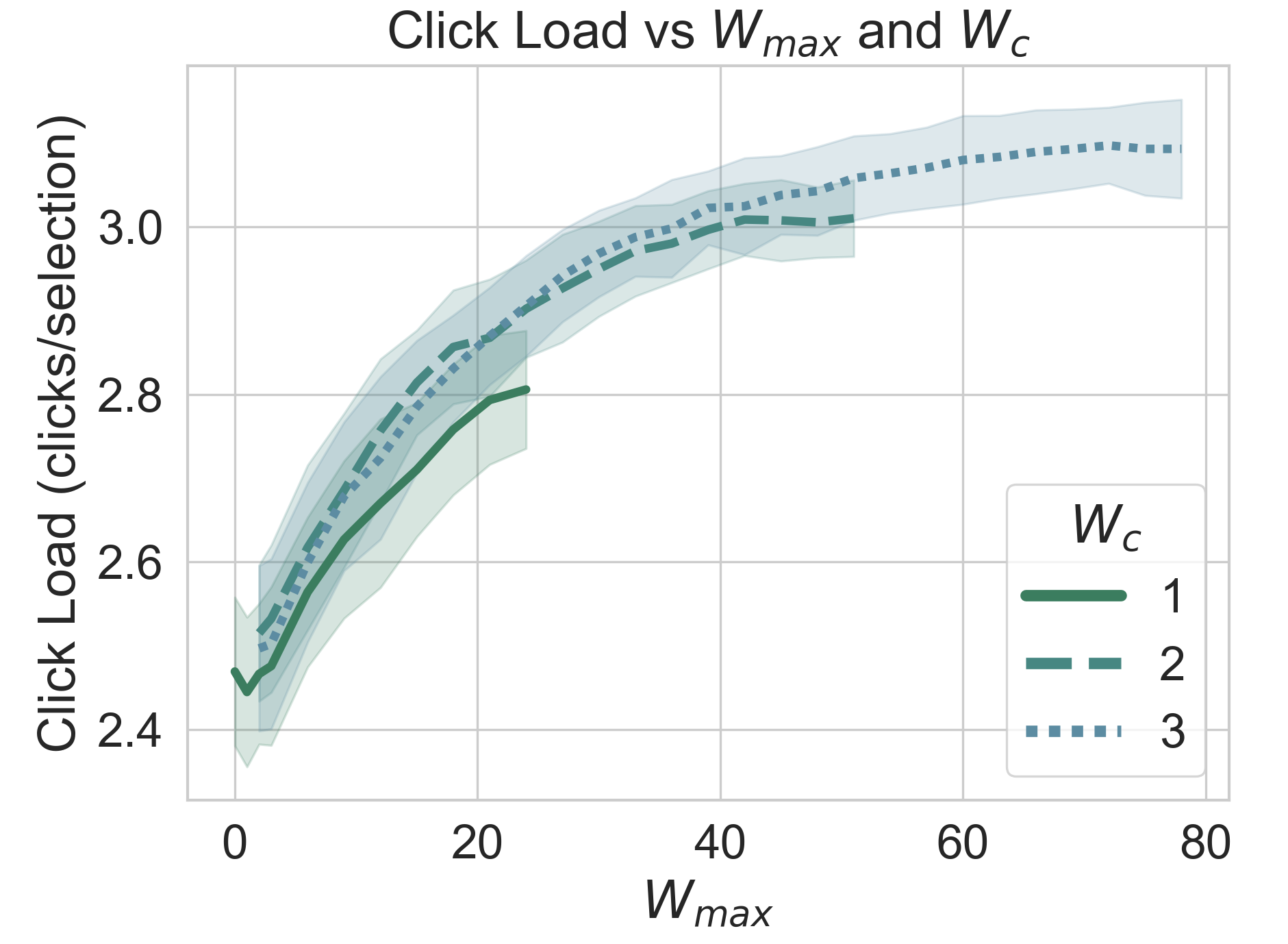}
    \caption{User model simulation results for entry rate (left) and click load (right) for the Nomon keyboard. We ran simulations across values of $W_c \in \{1, 2, 3\}$ (word completions per character) and $W_{\text{max}} \in \{1, 2, \ldots , W_c \cdot 26\}$ (total word completions displayed). Lines show the mean across 150 phrases. Error bands show bootstrapped 95\% confidence intervals. Samples were drawn from the click-time distribution of an experienced user --- hence the relatively high text-entry rate. Clicks per selection were, as expected, in line with experimental results for the picture selection task in our user study. Note that selections can include a word in this simulation. Participants averaged around 2.5 clicks/selection for the text entry task (with $W_{max}=17$) and reached 3.5 clicks/selection in the emoji task when there were 60+ options on the screen (which is similar to larger values of $W_{max}$).}
    \Description{(left) a plot with entry rate in characters per minute along the y-axis and $W_{\text{max}}$ ranging from integer values of 0 to 80 along the x-axis. Three separate lines plot the resulting entry rate as a function of $W_{\text{max}}$ for the 3 possible values of $W_{\text{c}}$. Each line quickly reaches its maximal entry rate value around $W_{\text{max}} = 5$ before leveling off. The line representing $W_{\text{c}}=3$ plateaus to the highest entry rate of the three. (right) a plot with click load in clicks per selection along the y-axis and $W_{\text{max}}$ ranging from integer values of 0 to 80 along the x-axis. Three separate lines plot the resulting entry rate as a function of $W_{\text{max}}$ for the 3 possible values of $W_{\text{c}}$. The three lines overlap significantly and follow a general trend of diminishing growth.}
    \label{figure:nomon_simulations}
    
\end{figure*}

We ran simulations of our user model to choose the number of word completions per character ($W_c \in \{1, 2, 3\}$) and the number of total word completions ($W_{\text{max}} \in \{1, 2, \ldots , W_c \cdot 26\}$) to display. Our simulations in Figure \ref{figure:nomon_simulations} showed $W_c = 3$ and $W_{\text{max}} = 17$ achieved the maximal text entry rate while keeping click load (number of switch activations per selection) to a minimum. More selection options can increase the click load depending on the width of the user's click-time distribution. The addition of word completions past 17 did not have a noticeable effect on entry rate gains; however, each additional word increased the click load. Therefore, we chose the lowest value of $W_{\text{max}}$ that achieved the maximal entry rate. As a result, there were at most 52 total options present on the screen at any one time.

Note that our results indicate that Nomon can handle more word completions than an RCS interface without seeing a drop in performance. We expected this behavior a priori: adding word completions increases the number of options RCS must cycle through before arriving at other options; Nomon is not so directly affected --- though greatly increasing the number of options in Nomon can ultimately require more clicks from the user to disambiguate. Further, our choice of fewer word predictions for the RCS interface was guided by our simulations and corroborated by existing literature on RCS optimization (detailed in Section \ref{subsec:rcs layout design}). However, no such work has determined optimal values for these parameters in Nomon. The original Nomon interface included up to $3 \cdot 26 = 78$ word completions \cite{nomonjournal} --- a number that our simulations suggest is too many.

\subsubsection{User-adjustable parameter.}
Users could set the rotation speed of the clocks $T$ to values of $T = 4e^{-l/10}$ seconds for $l \in \{0, 1 \ldots 20\}$. That is, smaller values of $l$ correspond to slower rotation, with $T \in [0.5, 4]$ seconds. Participants started with the slowest setting of $l=0$ and were allowed to increase or decrease $l$ by 1 between phrases.

\subsection{Open-Source and Browser-based Interface}
We originally designed our interfaces to run as standalone applications installed on a user's computer. We had started conducting an in-person lab study using these applications. But due to the COVID-19 pandemic, we suspended this study in March 2020. In the following months, we ported our applications to web interfaces running in a user's browser via HTML and JavaScript. Though it was a setback, we believe our implementations are now much more accessible than before; anyone can run our implementations in a standard browser without the need for local software installation. We encourage the reader to try it out at \url{https://nomon.app} and share any feedback. Our code for the Nomon application is open source and can be accessed via the link. At the same link, we also provide the code used to run the text-entry simulations with Nomon described in Section \ref{subsubsection: simulating a nomon user} and the simulations with RCS described in the supplemental materials.

\subsection{Word Predictions and Character Probabilities}
\label{subsection: Word Predictions and Character Probabilities}
In text entry applications, Nomon and RCS can make use of word predictions based on the text written so far. If the user has not started typing a word, Nomon and RCS predict the most likely words based on the previous words. If the user has started entering a word, the interfaces predict the most probable words that complete the current word. In calculating the clock phases, Nomon also uses the probability distribution over the next character given the current text.

To take advantage of this language information, our interfaces need to query language models at the word and character levels. Our language models were trained on data from a crawl of the web, social media, and movie subtitles. Our goal was to create models that approximate the sort of text AAC users might need for written or person-to-person communications. We started with 286\,B words of data and used cross-entropy difference selection \cite{moore_intelligent} to filter this down to 8.5\,B words of data. Filtering used a set of in-domain language models trained on conversational AAC-like data \cite{vertanen_aac_lm}, short email messages, and two-sided dialogues \cite{li2017dailydialog}. We trained our 12-gram character model using Witten--Bell smoothing \cite{bell_text_compression} and the 4-gram word model using modified Kneser--Ney smoothing \cite{chen_smoothing_tech}. The character and word language models had a compressed size of 782\,MB and 837\,MB respectively.\footnote{Specifically we used the large models from \url{https://imagineville.org/software/lm/dec19_char/} and \url{https://imagineville.org/software/lm/dec19/}.} We used BerkeleyLM \cite{pauls_berkeleylm} for efficient language model queries.

These language models have a large memory footprint and ranking words can be computationally expensive. Rather than performing these calculations in the browser, we instead built a web API that our interfaces queried. The Nomon and RCS interfaces predict the most likely words from the subset of a 100\,K vocabulary that matches the currently entered prefix. Specifically, for this subset of words, we calculated the log probability of the remaining letters of each word, including a trailing space, under the character language model. To each word, we also added the log probability of the word under the word language model. Both language models conditioned on any text written prior to any current partial word.

Since we could not control the latency between users and our language model server, we added caching API endpoints. These caching endpoints allowed the interface to look up all probabilities for all possible next selections by the user, thereby preventing noticeable lag after selecting a character or word. The language model was hosted as an API on a separate Apache Tomcat server with 8 CPUs and 8 GB of RAM. The predictions for a particular API call were computed in parallel to utilize all CPUs. Server load never exceeded 2 participants at a time to prevent lag in presenting predictions.

\subsection{A Single Switch via Webcam Input} \label{subsubsection:Apparatus and Software}

\paragraph{Motivation.} Individuals with severe motor impairments are challenging to recruit; their time and insight is particularly valuable. We feel ethically that we should thoroughly vet any system before consuming substantial time and effort of single-switch users. In line with this thinking, many preliminary studies on AAC methods include non--switch-using participants. In a survey of 42 studies on AAC software, 21\% included non--switch users \cite{koestersurvey}. Furthermore, when researchers include motor-impaired participants, the sample sizes can be limited. In the same survey, 21\% of studies included only a single participant \cite{koestersurvey}. Non-switch users provide a way to ameliorate noisy data from (often few or one) motor-impaired users and to gain statistical power to distinguish interface performance \cite{Higginbotham}. Since non--switch users play a prominent role in the study of AAC software, we want to ensure they approximate the target population as well as possible. For our studies, we designed a single switch based on input from a webcam. In particular, we chose our webcam switch over a button press in order to better approximate motor-impaired reaction times, as we describe below.

\begin{figure}[tb]
	\centering
	\includegraphics[width=\linewidth]{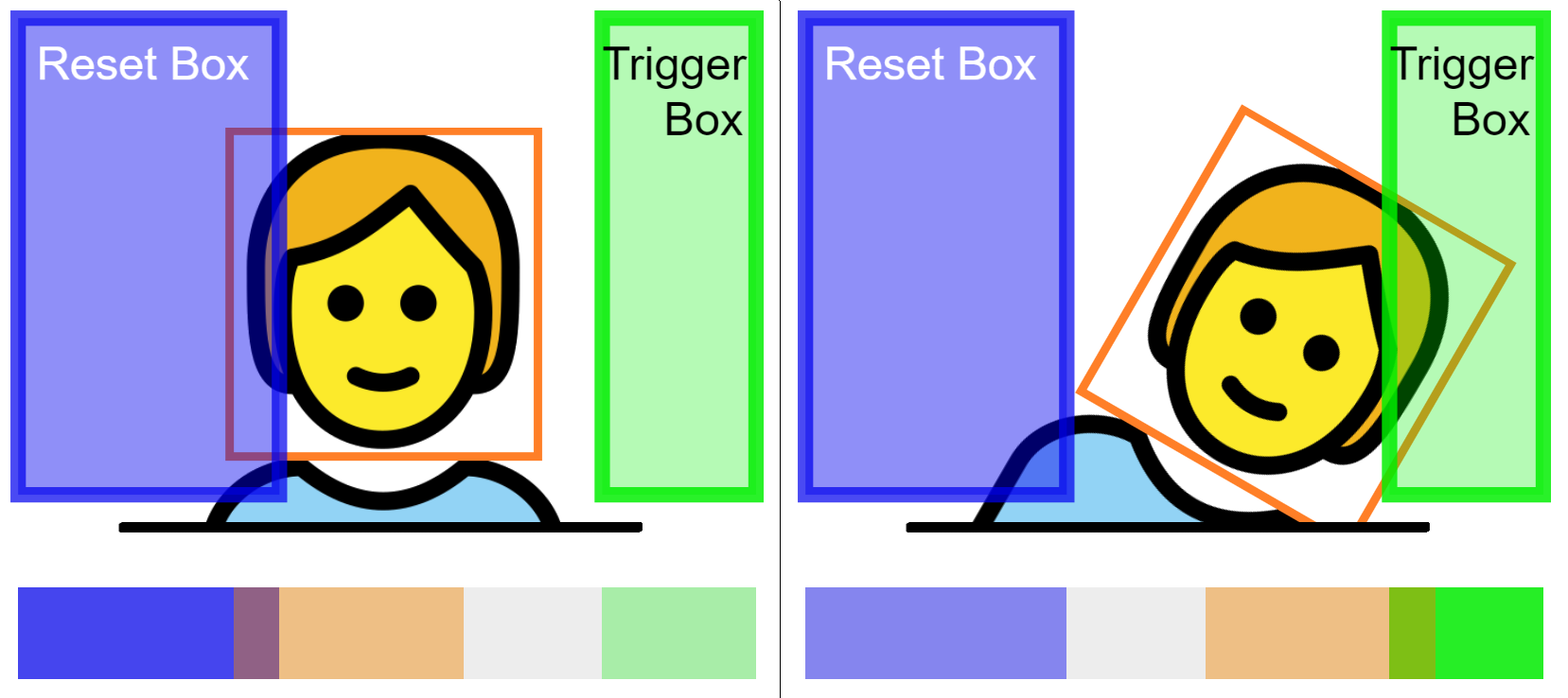}
	\caption{An illustration of our webcam switch, designed to approximate motor-impaired reaction times. To cause a switch click, the user first activates the blue reset box --- which is adjusted to their resting position --- and then activates the green trigger box --- which is adjusted so that they must move their head to the side. The emojis were adapted from \cite{personemoji}.}
	\Description{An illustration of a person using our webcam switch. The person has an orange box surrounding their head representing the detected location of their face. To the left  is a blue box representing the reset region, and to the right is a green box representing the trigger region. The image shows two positions side by side: first the resting position and then the trigger position. The resting position shows the person sitting upright with their face intersecting the blue region. The trigger position shows the person tilting their torso to the side with their face intersecting the green region. }
	\label{figure:Switch Viacam}
\end{figure}

\paragraph{Switch details.} Our webcam switch tracks the movement of a user's face and displays their current face location in the form of an orange box. Users activate the webcam switch by moving their head in and out of two regions in succession: (1) a reset region and (2) a trigger region (Figure \ref{figure:Switch Viacam}). To ``click'' (i.e., to activate the webcam switch) the user first moves their head so that the orange box intersects the blue reset region, and then moves their head into the green trigger region. The reset box activates when the participant is in a neutral position, and the trigger box activates when the participant moves their head to the other side. Triggering our webcam switch requires a wide motion of the user's torso and therefore decreases reaction time relative to a simple button press.

\paragraph{Preliminary justification.}
The first author made a limited comparison of reaction times using our webcam switch and a tactile button. We found the reaction times using our webcam switch were much closer to observed data of motor-impaired switch operation, collected in \citep{koesterpsycometric,koesterscanwizzard}. Using data from our own user study, we formally validate this finding; see Section \ref{section:approximating_motor_impairments}.


\section{User Study} \label{section:User Study}

Here we describe the two tasks in our user study: a text-entry task and a picture-selection task --- as well as results for both tasks. We chose an extended study design with 10 sessions to provide insight into experienced use of both interfaces. We conclude that experienced users of Nomon find both tasks faster and easier than row-column scanning.

\subsection{Participants}
We recruited 13 non--switch-using participants through emails sent to university and community mailing lists. All participants provided written, informed consent. Our experimental protocol was approved by our institutional review board. 8 participants were female and 5 were male. Their ages ranged from 19 to 76 (mean 35, sd 20). 8 were currently attending university, and their locations varied across the United States. None were familiar with either interface or with single-switch text entry software.

In addition, we recruited a single-switch user to trial the Nomon keyboard. They have an advanced form of spinal muscular atrophy and have over 14 years of experience using single-switch scanning. They use an EMG switch and EZ Keys row-column scanning for their daily computer interaction. This switch was used throughout their involvement in this study.

\subsection{Procedure}
\label{subsubsection:Study Procedure}

The non--switch participants took part in 10 sessions and paced themselves after the initial session. We instructed them to aim for 1–2 sessions per week, with no more than 1 session per day. Participants took around 8 weeks to finish the study.

In the first session, we explained the purpose of the study and obtained informed consent. We considered this session as practice since participants used both interfaces for less than 5 minutes each. We did not analyze results from this practice session. The first author was present via video conferencing during the first session and second session to introduce the study and answer any questions. 

Sessions 2--9 were structured as follows. Participants used the Nomon and RCS interfaces for 20 minutes each to perform the text-entry task described in Section \ref{subsection:Text Entry Task}. We alternated which interface (Nomon or RCS) each participant used first to achieve a near-even split. We had participants alternate which interface they used first from session to session. In the study, we referred to the two interfaces simply as A and B to minimize bias towards Nomon~\cite{dell_better}.

In sessions 2, 5, and 9, participants completed a questionnaire after using each interface. In sessions 2 and 9, participants also completed a NASA Task Load Index (TLX) \cite{nasatlx}. The NASA TLX aims to measure the ``load'' experienced by a user when performing a task. Sessions 2 and 9 included the sources-of-load section. In session 5, we administered only the magnitude-of-load section. In session 6, participants completed a reaction time task before using either interface; see Section \ref{section:approximating_motor_impairments} for full details.

In session 10, we had participants perform a picture selection task described in Section \ref{subsection:Picture Selection Task}. Participants used each interface for this task for 20 minutes, for a total of 40 minutes. After each method, we administered the NASA TLX (including the sources-of-load) as well as a questionnaire.

\subsection{Experiment 1: Text Entry Task}
\label{subsection:Text Entry Task}

\begin{figure}[tb]
	\centering
	\includegraphics[width=\linewidth]{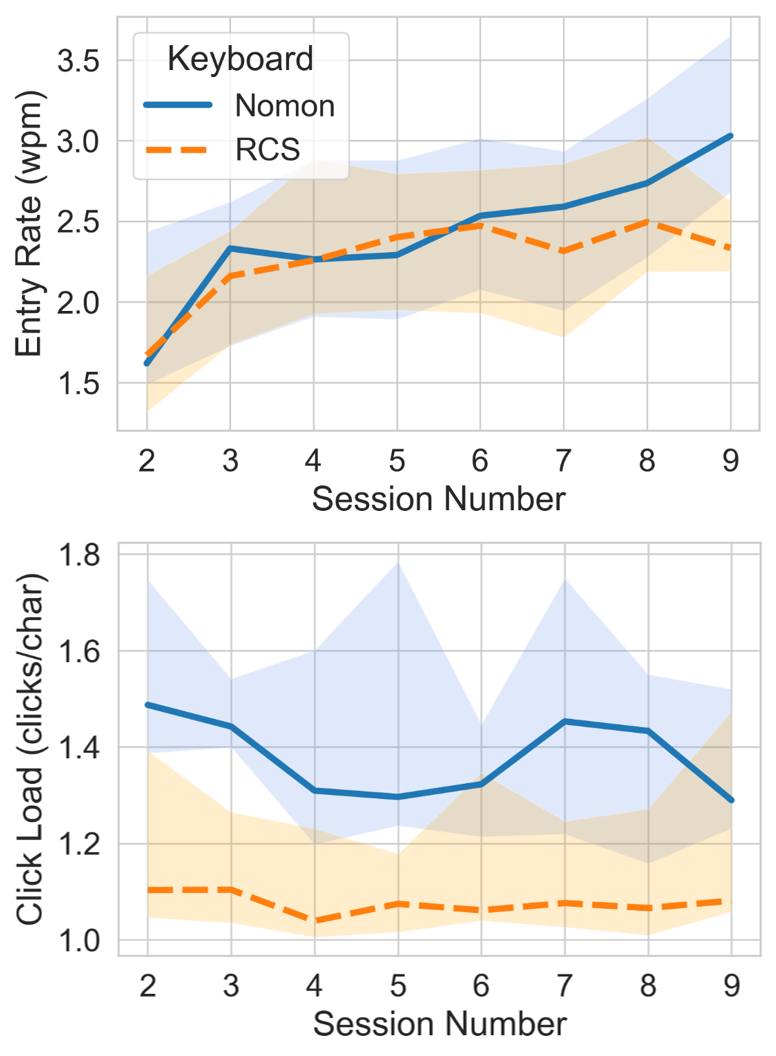}
	\caption{Median text entry rate (\emph{top}) and click load (\emph{bottom}) across all 8 sessions of the text entry task. Error bands show the first and third quartiles of the distributions. The upper curve corresponds to Nomon in both plots.}
	\Description{(top) A plot with entry rate in words per minute along the y-axis and session numbers ranging from 2 to 9 along the x-axis. Two lines are plotted representing Nomon and RCS. The RCS line plateaus after session 5, while the Nomon line continues to trend upward till the end. (bottom) A plot with clock load in clicks per selection along the y-axis and session numbers ranging from 2 to 9 along the x-axis. Two lines are plotted representing Nomon and RCS. Both lines generally trend horizontal with wide error bars. The RCS line is underneath the Nomon line.}
	\label{figure:Longitidunal Results}
\end{figure}

\subsubsection{Procedure}
\label{subsubsection:text entry proceedure}
In the text-entry task, participants typed as many phrases as possible in a 20-minute time period with each interface. Participants signaled that they were finished transcribing a phrase by pressing the ``Enter'' key. We drew phrases uniformly at random (without replacement, both within sessions and across sessions) from a set of phrases. Our aim was to choose phrases that were easy to remember and that represent text people might chose to write when not artificially constrained by AAC software. To those ends, we constructed two phrase subsets: (1) an out-of-vocabulary (OOV) phrase subset: a set of phrases containing exactly one word not in the language model (described in Section \ref{subsection: Word Predictions and Character Probabilities}) and (2) an in-vocabulary (IV) phrase set: a set of phrases for which all words were in our language model. We derived both phrase subsets from the ``challenging phrase set'' developed in \cite{velociwatch}. These phrases were all manually reviewed in \cite{velociwatch} to ensure that they were easy to remember. The IV and OOV subsets had a mean phrase length of 7.15 (sd 1.60)  and 7.24 (sd 1.64) words respectively.

Finally, we constructed our full phrase set by mixing the subsets at a ratio of two in-vocabulary phrases for every one out-of-vocabulary phrase. This mixture ensures we test both the word-completion and the general text-entry abilities of both interfaces. While word completions allow much faster text entry in single-switch text-entry systems \cite{Koester2014enhancing}, unusual words can sometimes arise, e.g.~individuals' names, places, and abbreviations. The previous study of text entry with Nomon \cite{nomonjournal} made use of the MacKenzie phrase set \cite{mackenzie_phrase}. This phrase set has been shown to have a low incidence of OOV words \cite{velociwatch}. Further, while the MacKenzie phrases may have contained some OOV words, the study did not explicitly examine the effects of these OOV words on text entry performance \cite{nomonthesis} --- whereas we are able to separately examine IV and OOV performance in the present study.

\begin{figure}[tb]
	\centering
	\includegraphics[width=\linewidth]{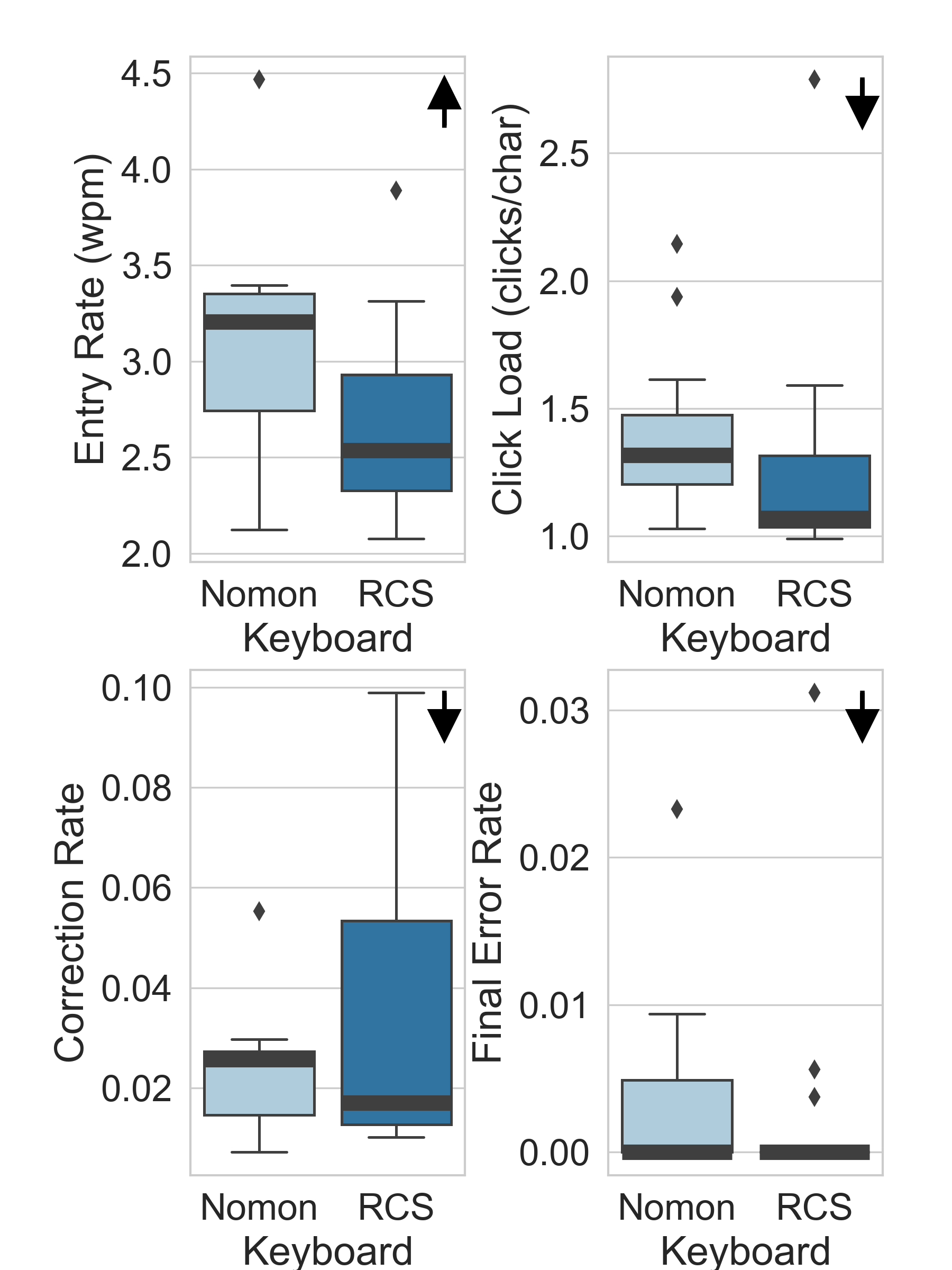}
	\caption{Metrics for sessions 8 and 9. The colored regions are the first to third quartiles of the distributions. The whiskers show the 5th and 95th percentiles of the distributions. Arrows in the top right show the direction of better performance.}
	\Description{Four box plots showing the mean values and distribution of participant's entry rates, click loads, correction rates, and final error rates. Mean and median values from these plots are provided in Table \ref{tab:text entry stats}.}
	\label{figure:Text Entry Task Results}
\end{figure}

\subsubsection{Performance metrics}
We calculate \textit{text-entry rate} in words per minute (wpm). We define a word as 5 characters including space. We include only characters present in the final output in our count (i.e., no corrected or undone text). We measured the time interval from the first switch input in each phrase up until the participant signaled they were finished with a phrase.

We define \textit{click load} as clicks per character (cpc) in the final output of a phrase (excluding corrected characters). Activating a switch is often an arduous task for individuals with severe motor impairments; therefore, it is important to consider this metric and not merely the text-entry rate when assessing effectiveness of a single-switch method.

We define \textit{correction rate} as the number of corrections divided by the total number of selections a user required to type a phrase. A correction is a selection of any of the Undo, Backspace, or Clear options. The correction rate gives a measure of how often a user made a mistake when typing.

We define \textit{final error rate} as the Levenshtein distance between the target phrase and a participant's final text output divided by the length of the target phrase. The Levenshtein distance measures how many character insertions, deletions, or substitutions are required to go from one string to another.

\subsubsection{Results}

\label{subsubsection:Text Entry Task Results}

\paragraph{Expert Performance}  We are interested primarily in comparing the performance of expert users; therefore, we restrict our analyses to data aggregated over the final two sessions (eight and nine). We performed a Shapiro\textendash Wilk test for normality in the paired samples across the two interfaces. We found the normality assumption was violated for click load $(W = 0.667,  p<0.001)$ and final error rate $(W = 0.479,  p<0.001)$. Where normality could be assumed, we used a dependent t-test (denoted as $t$); otherwise we used a Wilcoxon signed-rank test. Table \ref{tab:text entry stats} shows numerical results and the corresponding significance tests.

\begin{table*}[tb]
    \centering
    \begin{tabular}{lrlrllll}
    \toprule
        Metric  & \multicolumn{2}{c}{Nomon} & \multicolumn{2}{c}{RCS}   & \multicolumn{3}{c}{Statistical Test} \\
                & mean  & median            & mean & median             & \\
    \hline
        \textbf{Entry Rate (wpm)}       & 3.10 & 3.21     & 2.69 & 2.53       & $t(12) = 2.88$    & $r=0.639$ & \textbf{\emph{p} = 0.014}\\
        \textbf{Click Load (cpc)}       & 1.39 & 1.32    & 1.27 &  1.07      & Wilcoxon          & $r=0.187$ & \textbf{\emph{p} = 0.046}\\
        Correction Rate                 & 0.0215 & 0.0257     & 0.0354 & 0.0170        & $t(12) = -1.76$   & $r=0.452$ & $p = 0.104$\\
        Final Error Rate                & 0.0038 &  0.000     & 0.0031 & 0.000         & Wilcoxon          & $r=0.11$ & $p = 0.499$\\
    \hline
        \textbf{IV Entry Rate (wpm)}    & 3.48 & 3.37      & 3.07 & 2.92 & $t(12) = 2.25$    & $r=0.410$ & \textbf{\emph{p} = 0.033}\\
        \textbf{OOV Entry Rate (wpm)}   & 2.32 & 2.43      & 1.90 & 1.81 & $t(12) = 3.94$    & $r=0.620$ & 
        \textbf{\emph{p} < 0.001}\\
        IV Correction Rate              & 0.019 & 0.016    & 0.029 & 0.016 & $t(12) = 2.25$    & $r=0.319$ & $p = 0.174$\\
        \textbf{OOV Correction Rate}    & 0.026 & 0.022    & 0.051 & 0.035 & $t(12) = -2.41$    & $r=0.435$ & 
        \textbf{\emph{p} = 0.023}\\
    \hline
        NASA TLX, session 2                   & 38.3 & 41.4    & 35.3 & 34.6      & $t(12) = 0.65$    & $r=0.161$ & $p = 0.525$\\ 
        NASA TLX, session 5                   & 32.9 & 33.4    & 32.5 &  33.9     & $t(12) = 0.65$    & $r=0.160$ & $p = 0.905$\\
        \textbf{NASA TLX, session 9}          & 27.0 & 27.0    & 33.4 &  33.4     & $t(12) = 0.12$    & $r=0.575$ & \textbf{\emph{p} = 0.032} \\
    \bottomrule
    \end{tabular}
    \caption{Mean and median result values and statistical tests for the text-entry task in the user study. Results are for sessions 8 and 9. Metrics in bold were significant.}
    \label{tab:text entry stats}
\end{table*}

Figure \ref{figure:Text Entry Task Results} displays the aggregate text entry metrics for sessions 8 and 9, for all participants. Participants typed 1.15 times faster using Nomon over RCS; however, they had a slightly higher click load using Nomon compared to RCS. The first published Nomon study~\cite{nomonjournal} found that participants typed 1.35 times faster using Nomon over RCS. The discrepancy in results might be attributed to the noise we have introduced via the webcam switch, as we observed larger error bars compared to~\cite{nomonjournal}. 
As in Figure~4 in \cite{nomonjournal}, the RCS entry rate here seemed to plateau before the entry rate of Nomon. In our study, the RCS plateau is reached in a later session, which might be expected due to the learning curve associated with a more noisy switch. We found no significant difference in correction rates or final error rates between the interfaces. 

\begin{figure}[tb]
    \centering
    \includegraphics[width=.8\linewidth]{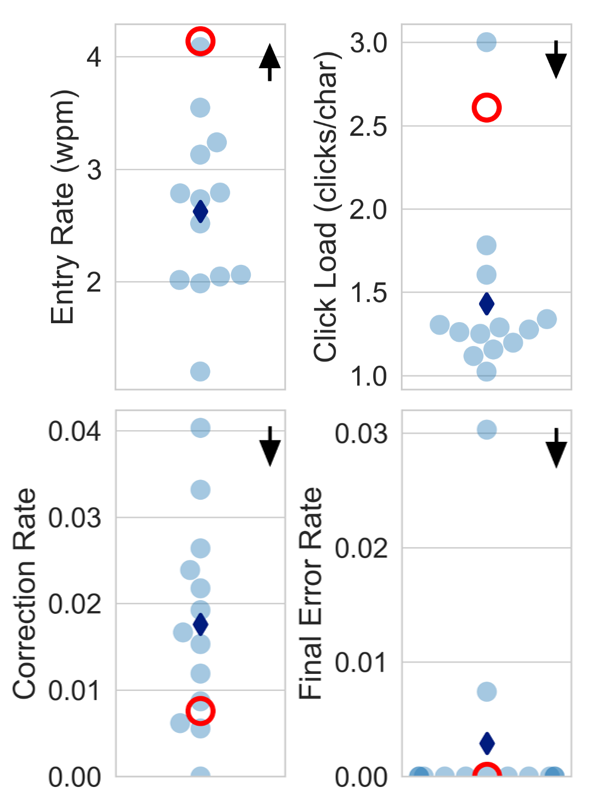}
    \caption{Comparison of text entry metrics for the non--switch users and the motor-impaired user after 80 minutes of prior practice (equivalent to session 6 for the non--switch users). Each  filled light-blue circle represents a single non--switch user, and the population mean is given by a filled dark-blue diamond. Red circles represent the motor impaired user. Arrows in the top right show the direction of better performance; e.g.\ we prefer a higher text-entry rate.}
    \Description{Four plots showing the mean value of the single-switch user's entry rate, click load, correction rate, and final error rate. The switch-user's values are plotted atop the distribution of the non--switch-user participants' data. The switch user had a higher entry rate than all the non--switch users and a higher click load than all but one non--switch user. The switch-user's correction rate was lower than most of the non--switch users, and their final error rate was near zero along with most other participants.}
    \label{fig:mi_final_metrics}
\end{figure}

\paragraph{Switch User Performance}
We recruited a single-switch user to complete the text entry task using the Nomon interface. We do not compare their performance between Nomon and RCS directly, as they use an RCS system daily and it would not lend a fair comparison. Rather, we compare their performance with Nomon to that of the hindered, non--switch users.

The participant regularly uses the RCS software EZ Keys with a $100$ millisecond scan speed. They have abbreviation expansion and custom, task-specific word completions to speed text entry. Utilizing this optimized setup, they have self-reported to type at an impressive $13$ wpm. We note the fast scan speed at which this switch user regularly uses an RCS interface. The switch user's proficiency with their switch allowed them to use Nomon with a rotation period of 0.76 seconds --- a considerably faster period than the average 3.35 seconds of the non--switch-using participants in this study. While this level of switch accuracy and speed may not be representative of a majority of single-switch users, this particular switch user's proficiency and associated quick communication speed was why we felt comfortable having this user pilot test our study methods. The switch user has provided us with insights into our study and software design that will prove invaluable in our following work with more diverse members of the target population.

We show the switch user's results alongside those of the non--switch users from session 6 (after an equivalent 80 minutes of practice) in Figure \ref{fig:mi_final_metrics}. The switch user's sessions ran identically to the text-entry-task sessions for the non--switch-users. The correction rate and final error rate of the switch user both fell within those of the non--switch users. However, the switch user had a considerably higher text entry rate (1.5 times faster; 4.14 wpm) and click load (1.8 times larger; 2.61 cpc). The switch user's shorter rotation period (4.4 times faster) than the non--switch users may account for this increase in both entry rate and click load. While a shorter rotation period may have allowed the switch user to minimize dead-time and thus increase their entry rate, the shorter period may have caused them to be less precise and require more clicks per selection.

\begin{figure}[tb]
    \centering
    \includegraphics[width=.9\linewidth]{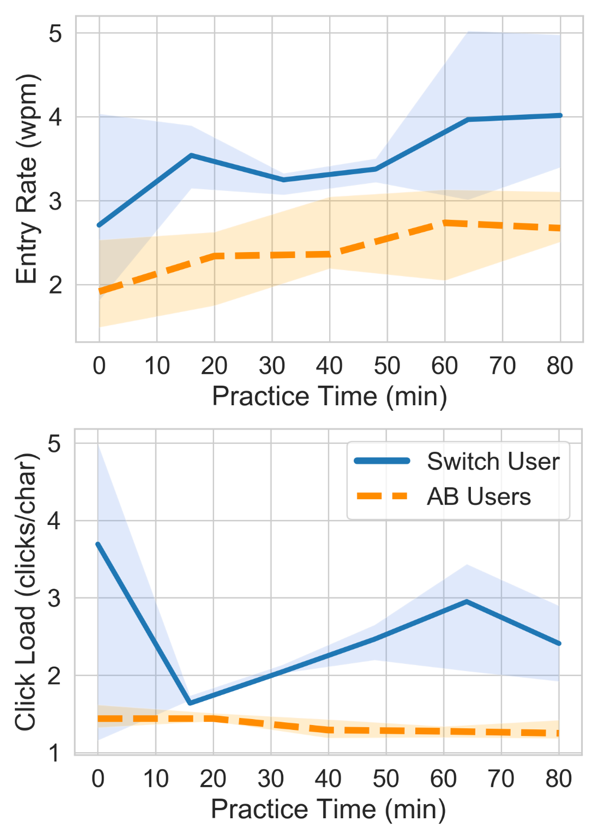}
    \caption{Median text entry rate (\emph{top}) and click load (\emph{bottom}) for the switch user (\emph{blue, solid line}) and hindered, non--switch users (\emph{orange, dashed line}) in Nomon. The x-axis shows how long users practiced with Nomon. At any time point, we plot a summary of the switch user's distribution over performance on individual phrases, while we plot a summary of the non--switch users' performances across the 13 participants. The error bands show the first and third quartiles of the relevant distribution.}
    \Description{(top) A plot with entry rate in words per minute along the y-axis and practice time ranging from 0 to 80 minutes along the x-axis. Two lines are plotted representing the switch-user and the median of the non--switch-user participants. Both lines trend upward till the end, with the switch-user consistently having a higher entry rate than the non-switch users. (bottom) A plot with clock load in clicks per selection along the y-axis and practice time ranging from 0 to 80 minutes along the x-axis. Two lines are plotted representing the switch-user and the median of the non--switch-user participants. The non--switch users consistently average a lower click load than the switch-user. The switch-user's click load does not appear to settle and has a wide variance.}
    \label{fig:mi_learning_curve}
\end{figure}

Further, we compare the learning curves of the switch user and the non--switch users with Nomon in Figure \ref{fig:mi_learning_curve}. The switch user had a consistently higher entry rate compared to the non--switch users at identical practice times with Nomon. The switch user's performance also increased with practice, much like the non--switch-using participants. However, the click load of the switch user varied much more throughout their practice sessions. After around 20 minutes of practice, the switch user reached their lowest click load of 1.6 cpc. The click load then continued to increase throughout the remaining sessions. This minimal click load occurred when the switch user had a rotation period of 1.62 seconds, with the larger click loads occurring as the user progressively shortened the rotation period.

\begin{figure}[tb]
    \centering
    \includegraphics[width=\linewidth]{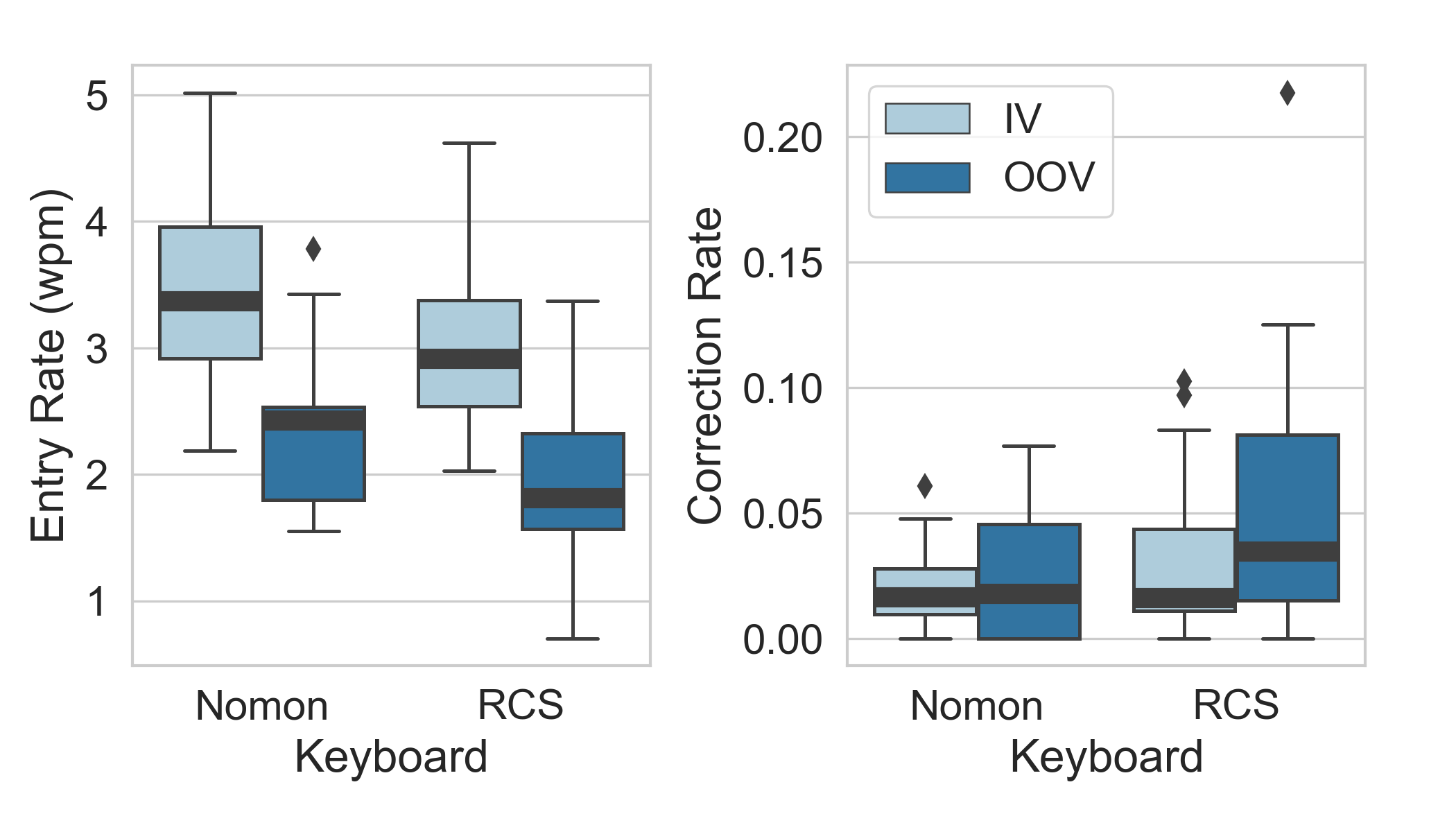}
    \caption{Entry rates and error rates for Nomon and RCS for in-vocabulary (IV) and out-of-vocabulary (OOV) phrases. In each paired comparison, IV appears to the left of OOV. Results are from sessions 8 and 9. The colored regions are the first to third quartiles of the distributions. The whiskers show the 5th and 95th percentiles of the distributions.}
    \Description{two box plots showing the mean values and distribution of entry rates and correction rates on in-vocabulary and out-of-vocabulary phrases. Mean and median values from these plots are provided in Table \ref{tab:text entry stats}.}
    \label{fig:oov_rates}
\end{figure}

\paragraph{Challenging Text Entry}
The combination of IV and OOV phrases allows us to test both the word completion and general text entry abilities of the interfaces. As evident in Figure \ref{fig:oov_rates}, we found that the addition of a single OOV word in a phrase can considerably lower text entry rates in both interfaces. This result is consistent with work investigating the effect of OOV words in mobile text entry in \cite{velociwatch}. Users were able to better handle these OOV words using Nomon. They typed OOV phrases 1.22 times faster and with with half as many corrections using Nomon over RCS. This difference suggests Nomon may be better suited to less predictable text composition than RCS. Indeed, Nomon's probabilistic selection mechanism does not seem to favor word completions for quick selection as dramatically as RCS (which dedicates the first scan row to word predictions that are useless for OOV words). Furthermore, users also performed better with Nomon on IV phrases, though to a lesser extent; they typed 1.13 times faster using Nomon, but had no significant difference in correction rate.

\begin{figure}[tb]
	\centering
	\includegraphics[width=.8\linewidth]{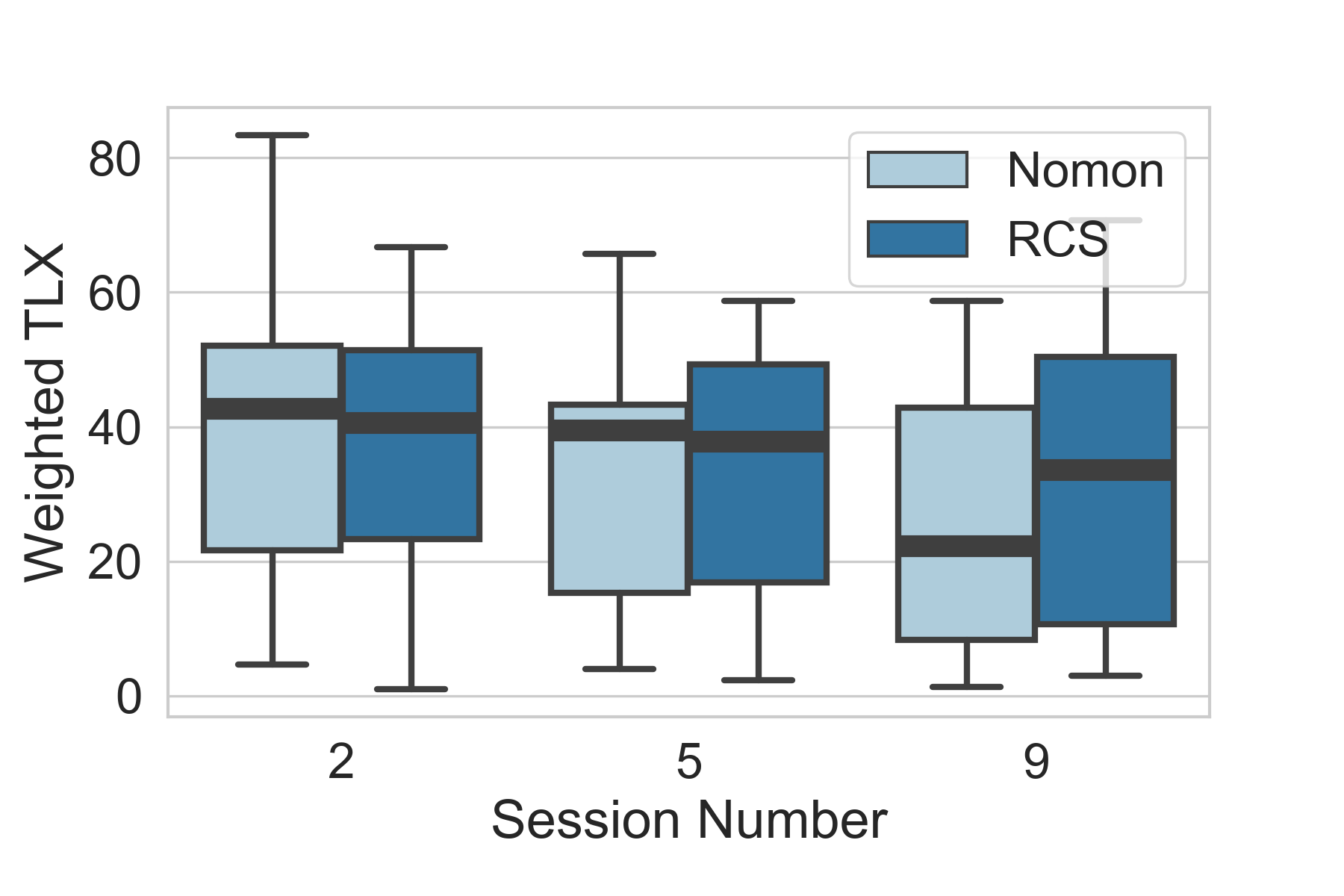}
	\caption{Results of the NASA Task Load Index administered following session 2, 5, and 9. In each paired comparison (per session), Nomon appears to the left of RCS.}
	\Description{a box plot showing the mean values and distribution of weighted TLX scores in sessions 2, 5 and 9. Mean and median values from these plots are provided in Table \ref{tab:text entry stats}.}
	\label{figure:TLX}
\end{figure}

\subsubsection{Subjective Feedback}
\label{subsubsection:Subjective Ratings}
\begin{figure*}[tb]
	\centering
	\includegraphics[width=.9\linewidth]{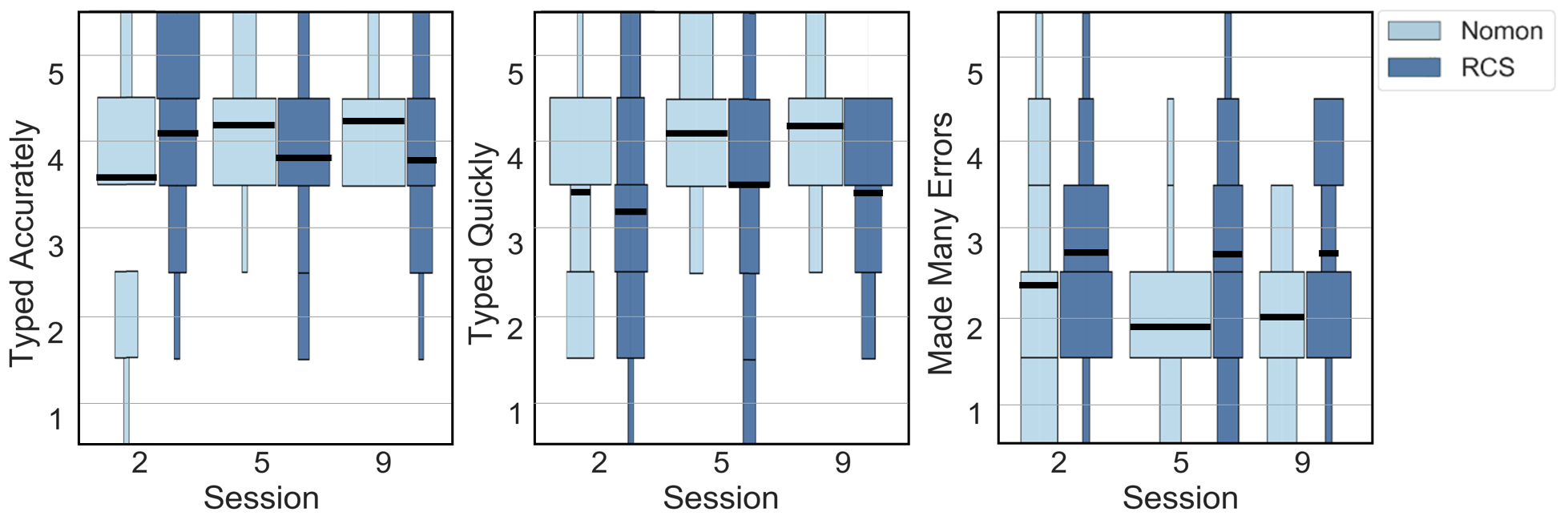}
	\caption{Plot of the results of questionnaires administered in the beginning, middle, and end of the user study. The relative width of the color regions denote how many participants gave a statement that Likert score. Statements were presented in the form ``In this part of the study, I felt that I typed quickly.'' Participants responded on a scale from 1 (strongly disagree) to 5 (strongly agree). Means are represented by horizontal black lines. In each paired comparison (per session), RCS appears to the right of Nomon. 
	}
	\Description{   }
	\label{figure:Survey}
\end{figure*}

We assessed user experience with questionnaires for each interface in the second, fifth, and ninth sessions. Participants indicated their agreement with a series of statements on a scale from 1 to 5, with 1 indicating ``strongly disagree'' and 5 indicating ``strongly agree.'' The distribution of responses across the sessions appears in Figure \ref{figure:Survey}.

As evident in Figure \ref{figure:Survey}, participants increasingly felt they typed faster, more accurately, and with fewer errors as they used Nomon more. Conversely, participants generally rated RCS the same in these three areas throughout the study. Figure \ref{figure:TLX} shows no notable difference in the overall NASA TLX scores between RCS and Nomon in sessions two and five. However, in the final session, participants rated Nomon as having a lower task load $(t(12)=0.12, p<0.032)$. This result further indicates that participants increasingly found Nomon easier to use with practice.

At the conclusion of the text entry task, we asked participants to choose between the two interfaces. 12 out of 13 participants indicated that they preferred typing with Nomon over RCS. Common reasons for this choice were that Nomon is ``more forgiving with errors,'' there is ``more flexibility'' and ``agency'' in the selection process, and ``less downtime waiting for scanning.'' 

We also received feedback from the switch user on their experience using Nomon. They noted, ``I observed more word predictions showing up as choices. This is where I see some real potential for increased typing rate (in terms of words per minute). Nomon is distinctly different from traditional scanning and may offer an easier path to higher text entry rates.'' The full responses from our participants can be found in the supplemental materials.

\subsection{Experiment 2: Picture Selection Task}
\label{subsection:Picture Selection Task}
\begin{figure}[tb]
	\centering
	\includegraphics[width=.8\linewidth]{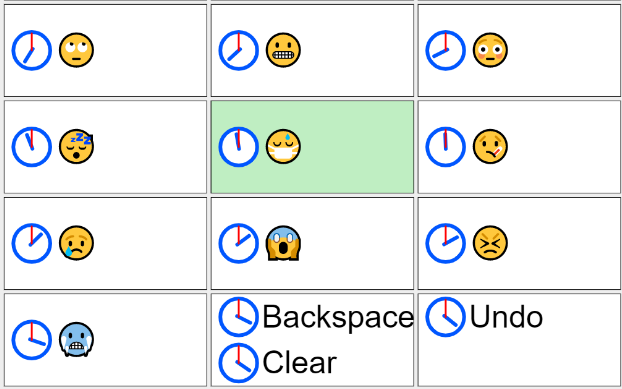}
	\caption{A portion of the Nomon interface for the picture-selection task. We adapted the RCS interface for the picture-selection task in a similar way.}
	\Description{A zoomed in section of the Nomon keyboard. A grid of emojis with Nomon clocks positioned to their left. One emoji's grid box is highlighted in a light green color.}
	\label{figure:Nomon Keyboard with Emojis}
\end{figure}

Text entry is a particularly important task for AAC users, so our user study focused on this task for most sessions. But there are many tasks of interest beyond text entry. Nomon has the advantage over RCS of being adaptable to tasks that need not fit into a grid. However, there exist tasks beyond text entry for which the two interfaces can be compared. In particular, when users choose among a large set of files on their computer, photos on a photo-sharing website, or products at an online vendor, these items can be arranged in a grid. Our aim was to encapsulate such a task and compare Nomon and RCS. We chose emojis as our set of options since we thought they would be easily recognizable by users and engaging for our participants.

For this experiment, we adapted the Nomon and RCS interfaces to include 60 emojis (Figure \ref{figure:Nomon Keyboard with Emojis}). The core functionality behind both interfaces remained the same. The interfaces highlighted the current target to avoid participants spending time searching through the options. This search time varies widely depending on how quickly a participant can find the next target; therefore including it in entry-rate calculations would introduce unnecessary variance. We chose 60 emojis because 60 was close to the maximum number of objects that could fit on the screens of both interfaces.

We expect Nomon to excel at this task. Under an uninformed prior (as in this task), previous work has shown that the number of switch clicks required to select a target in Nomon scales logarithmically with the number of options \cite{nomonthesis}. With a constant rotation speed, the time required for selection (excluding reaction time and the time spent searching for the desired option) should scale similarly.
By contrast, the mean number of scans to select an option in an RCS interface scales with the square root of the number of options $n$; the user must make an average of $\sqrt{n}/2$ row scans and then $\sqrt{n}/2$ column scans (if options are arranged in a square grid).

\subsubsection{Procedure and Performance Metrics}

We used the final session to test this alternative task. We expected users would have ample experience with both interfaces by the final session and therefore would not require multiple sessions to adjust to the picture-selection task. In lieu of English phrases, we asked participants to write sequences of five emojis at a time. We computed four metrics: entry rate (selections per minute), click load (clicks per selection), correction rate, and final error rate.

\begin{table*}[tb]
    \centering
    \begin{tabular}{lrlrllll}
    \toprule
        Metric  & \multicolumn{2}{c}{Nomon} & \multicolumn{2}{c}{RCS}   & \multicolumn{3}{c}{Statistical Test} \\
            & mean  & median            & mean & median             & \\
    \hline
        \textbf{Entry Rate (selections/min)}        & 6.64 & 6.65      & 4.88 & 4.59      & $t(12) = 5.24$    & $r=0.834$ & \textbf{\emph{p} < 0.001}\\
        \textbf{Click Load (cicks/selection)}       & 3.50 & 3.25      & 2.22 & 2.23      & $t(12) = 7.73$    & $r=0.912$ & \textbf{\emph{p} < 0.001}\\
        \textbf{Correction Rate}                    & 0.011 & 0.0095   & 0.026 & 0.0238     & $t(12) = -2.45$   & $r=0.577$ & \textbf{\emph{p} = 0.031} \\
        Final Error Rate                            & 0.0025 & 0.000   & 0.0043 & 0.000        & Wilcoxon      & $r=0.072$ & $p = 0.893$\\
    \hline
        NASA TLX, session 10                        & 27.2 & 27.1     & 27.7 & 27.7      & $t(12) = -0.22$   & $r=0.064$ & $p = 0.893$\\
    \bottomrule
    \end{tabular}
    \caption{Mean and median result values and statistical tests for the picture-selection task. Metrics in bold were significant.}
    \label{tab:picture entry stats}
\end{table*}

\begin{figure}[tb]
	\centering
	\includegraphics[width=.9\linewidth]{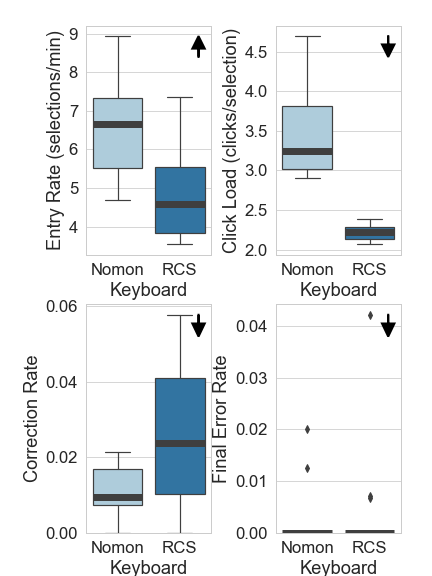}
	\caption{Performance metrics for the picture selection task in session 10. The colored regions are the first to third quartiles of the distributions. The whiskers show the 5th and 95th percentiles of the distributions. Arrows in the top right show the direction of better performance; e.g., we prefer a higher entry rate.}
	\Description{Four box plots showing the mean values and distribution of participant's entry rates, click loads, correction rates, and final error rates. Mean and median values from these plots are provided in Table \ref{tab:picture entry stats}}
	\label{figure:Emoji Task Results}
\end{figure}

\subsubsection{Results}
\label{subsubsection:{Picture Selection Task Results}}
A Shapiro\textendash Wilk test for normality in the paired samples found this assumption was violated for final error rate $(W = 0.479,  p<0.001)$. We used a dependent t-test (denoted as $t$) where normality could be assumed; otherwise we used a Wilcoxon signed-rank test. Table \ref{tab:picture entry stats} shows numerical results and the corresponding significance tests.

Figure \ref{figure:Emoji Task Results} shows user performance in the picture selection task in session 10. 
The benefits of Nomon were even more pronounced in picture selection compared to text entry. Participants selected targets substantially and significantly faster using Nomon --- an average of 36\% faster. This increase in entry rate comes with a trade-off in click load. Participants had a higher click load of 3.50 clicks per selection using Nomon, compared to 2.23 clicks per selection using RCS. However, we expected this increase given the conjectured logarithmic scaling in the number of required switch clicks \cite{nomonthesis}. Participants also made fewer corrections per selection using Nomon --- $1.1\%$ with Nomon versus $2.6\%$ with RCS. We found no significant difference in final error rates between the interfaces.
	
\section{Reaction Time Study}
\label{section:approximating_motor_impairments}

In Section \ref{subsubsection:Apparatus and Software}, we described the webcam switch we employ in our user study. 
In this section, we validate our claim that this switch yields a useful approximation of motor-impaired single-switch reaction times with non--switch user inputs.

\paragraph{Quantities to approximate.} There are two key quantities \citep{reactiontime} for single-switch operation that we aim to approximate:
\begin{itemize}
    \item Simple reaction time (SRT) --- SRT is the time difference between the introduction of a stimulus to a user and their subsequent response.
    \item Double click time (DCT) --- DCT is the amount of time between a user's successive switch activations. DCT measures how quickly a user can click their switch again after they have just clicked it.
\end{itemize}

SRT and DCT dictate how quickly users can operate single-switch software. E.g., if the scan delay or rotation time is too fast compared to a user's typical SRT, they may find the software unusable \cite{reactiontime}. RCS requires users to click their switch twice in immediate succession to select targets in the first column; if the scan delay is too fast compared to a user's typical DCT, they will be unable to select these targets.



\paragraph{Single-switch user and non--switch user data.} Dr. Heidi Koester graciously provided data on the SRTs and DCTs of non--switch-using and single-switch-using individuals that she and her colleagues collected --- namely, 10 motor-impaired users in \cite{koesterscanwizzard} and 10 motor-impaired users and 8 non--switch users in \cite{Koester2014enhancing}. While this data may not fully represent the diversity of motor-impaired switch users, it provides insight into the extent to which unhindered, non--switch-using participants can be unrepresentative of the motor-impaired population. Further, the data shows that by hindering non--switch users with our webcam switch method, we can better represent some subset of the motor-impaired population in two key metrics related to single-switch use (SRT and DCT). 

\label{subsection:Webcam Study Results}
\begin{figure*}[tb]
	\centering
	\includegraphics[width=150mm]{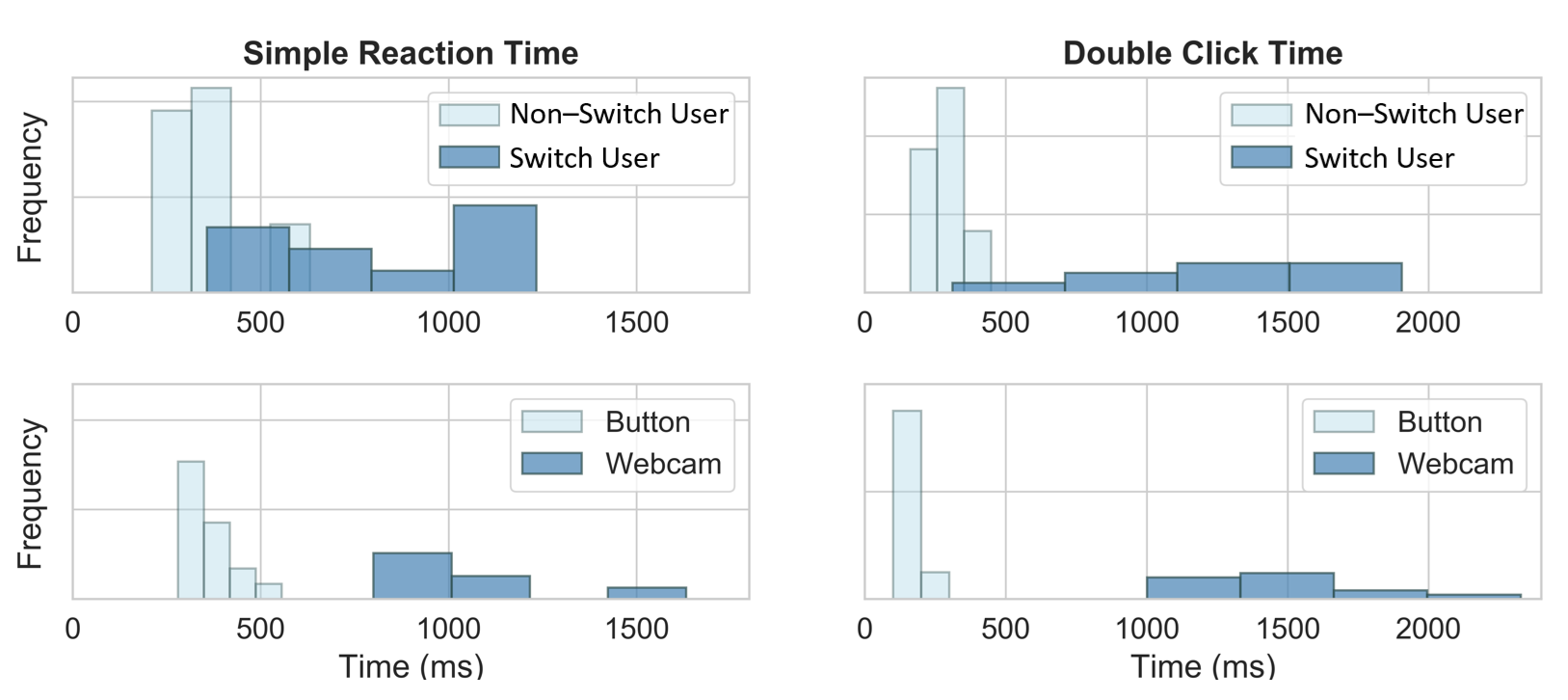}
	\caption{Comparison of the SRTs and DCTs of the two switch methods to those of the non--switch-using and single-switch-using populations. On the top row is the data from \cite{koesterscanwizzard, koesterpsycometric}, with the non--switch users in light blue (appearing left in each plot) and the single-switch users in dark blue (appearing right in each plot). The bottom row contains the data we collected with the button and webcam switches from our non--switch-using participants. The button histogram is in light blue (appears left in each plot), and the webcam histogram is in dark blue (appears right in each plot). SRTs are in the left column and DCTs in the right. Each data point is the average value across switch clicks of a particular participant using a particular switch method; we show a histogram over these data points.}
	\Description{(left) two sets of histograms of SRT stacked vertically with aligned x-axes. The top set shows the SRT distributions of the non--switch-user and switch-user populations. The bottom set shows the SRT distributions of the non--switch-user participants using a button and our webcam switch. The histograms of the non--switch-users and the button are well aligned to the left (around 500ms). The histograms of the switch-users and the webcam-switch are well aligned to the right (around 1000 ms). (right) two sets of histograms of DCT stacked vertically with aligned x-axes. The top set shows the DCT distributions of the non--switch-user and switch-user populations. The bottom set shows the DCT distributions of the non--switch-user participants using a button and our webcam switch. The histograms of the non--switch-users and the button are well aligned to the left (around 250 ms). The histograms of the switch-users and the webcam-switch are well aligned to the right (around 1500 ms).  }
	\label{figure:Reaction Times Plot}
\end{figure*}

\paragraph{Procedure.} We collected our data as an additional task added before the start of the sixth session of our user study. Participants used a web interface that first had them use our webcam switch and, secondly, their keyboard spacebar as a switch. Following \cite{koesterscanwizzard}, for each switch, we had the screen flash 30 different times at random intervals. We instructed participants to click their switch twice in quick succession after they saw the screen flash. 
For each switch method, we recorded 30 trials and calculated the participant's average SRT and DCT. These averages are visualized in the histograms in the bottom row of Figure \ref{figure:Reaction Times Plot}. 



\subsection{Results}

In the top row of Figure \ref{figure:Reaction Times Plot}, we see that single-switch users with severe motor impairments generally have an SRT and DCT much longer than non--switch users. There is also a wide variance among the motor-impaired population, with some individuals much faster than the mean, and some much slower.

Figure \ref{figure:Reaction Times Plot} shows that our webcam switch yields SRT and DCT values that are considerably more in line with those of the motor-impaired target population --- as compared to a spacebar switch. The webcam switch lowers the SRT of the participants from 350 ms (with the spacebar) to 1050 ms. By comparison, the mean SRTs for the non--switch using and single-switch using populations are 350 ms and 820 ms, respectively. Similarly, the webcam switch lowers the participants' mean DCT from 180 ms to 1400 ms. These DCTs are consistent with those from the non--switch using and single-switch using populations of 290 ms and 1460 ms. We conclude that our webcam switch technique substantially lowered both SRT and DCT to levels consistent with data from single-switch users with motor-impairments.

	\section{Discussion} \label{section:Discussion}

We investigated the effectiveness of Nomon as a method of single-switch communication. We evaluated the performance of Nomon over multiple sessions compared to the widely used row-column scanning method. In a text-entry task, participants typed 15\% faster using Nomon. However, they experienced a 10\% higher click load with Nomon. This higher click load could be problematic for users where switch activation is tiring. 

We are exploring ideas to mitigate this higher click load. One such idea is to use information from an eye gaze tracker, as users will undoubtedly be gazing towards the clock they are trying to select. Interestingly, the switch user who trialed Nomon commented that they ``notice[d] a sense of direct selection [with Nomon] (though technically it is not) akin to eye gaze interfaces. One important difference is that I did NOT experience the same eye strain/fatigue often associated with eye gaze mouse pointer navigation.'' 

Separately, we posit that it may be possible to allow just one click per letter for predictable words. Currently Nomon requires each individual character to pass a probability threshold before committing to that character. We believe we could postpone committing to any text until the end of a word (similar to how auto-correction works on a touchscreen keyboard). With only a noisy switch as input, designing how users signal the end of a word, correct errors, and enter difficult words would be challenging --- but would constitute interesting future work.

Participants continued improving with Nomon even in the final session, while they appear to plateau with RCS after session 5. Furthermore, participants found typing easier and faster using Nomon in the final sessions. 12 out of 13 participants indicated that Nomon was their preferred method of text entry. We had hoped eight text-entry sessions would be enough for Nomon performance to plateau, but users continued to improve even in our final session. Our results suggests a longer study may be necessary to fully explore Nomon's potential, especially when evaluating with motor-impaired users. 

To our knowledge, our study is the first to investigate single-switch input of text containing difficult out-of-vocabulary (OOV) words. When selecting OOV words, the word language model is not active but the character language model still provides a non-trivial prior over common sequences of characters.  We found Nomon significantly reduced the need to perform corrections and significantly increased entry rate on OOV phrases. This advantage is important since error correction can be a frustrating process, especially using a single switch. Our interfaces limited word predictions to a vocabulary of 100\,K words. We think further improvements in Nomon's efficacy for OOV words may be possible by expanding the prediction engine's vocabulary to a larger word list when the set of predictions becomes sparse or empty. This word list could be created from timely online data sources (e.g., Twitter) and predictions ranked via a language model with a subword vocabulary and trained on enormous amounts of data (e.g., GPT-2 \cite{radford_gpt2}). Our participants also suggested other improvements such as increasing the probability of the undo clock, and removing word predictions that were not selected to free up space for other words.

We explored applications beyond text entry with a picture selection task. The picture-selection task gives the user a large number of options with a uniform prior. Here, the benefits of Nomon were more pronounced as participants selected options 35\% faster and with 63\% fewer errors. On the other hand, participants had a 53\% higher click load; this increase in click load seems to be fundamental to Nomon's flexible selection scheme, where the number of switch clicks required for selection should scale logarithmically with the number of options. These results are promising for future work using Nomon in applications beyond text entry. In particular, it would be interesting to explore tasks that can leverage a prior over targets learned from individual users (e.g., the sequence of links clicked in an application or the control of home IoT devices). 

To aid our studies above, we designed and validated a webcam-based switch technique for better approximating motor-impaired operation of a single switch with non--switch-using participants. We found that the simple reaction times (SRTs) and double click times (DCTs) of non--switch users with a physical button were unrepresentative of SRTs and DCTs of single-switch users with motor impairments. Our webcam method artificially lowers a non--switch user's reaction times to more closely resemble the reaction times of single-switch users with motor impairments. Using this technique with our participants allowed us to collect data that more closely resembles that of our target population while
recruiting non--switch users as participants.

To further evaluate Nomon, we are planning a similar user study to the one reported here but with a group of motor-impaired users. 

\section{Conclusion}
\label{section:conclusion}

To conclude, we made the Nomon interface more accessible through collaboration with switch users and AAC specialists. We further optimized the design of the Nomon interface via computational simulations. We developed a webcam-based technique to simulate the click timing of motor-impaired users. Our user study results alongside our initial trial with a switch user show that Nomon may currently provide accelerated text input for single-switch AAC users. In their final session (after 2.5 hours of practice), users wrote 15\% faster using Nomon than with conventional row-column scanning. We found this speedup was even more pronounced when composing challenging text containing out-of-vocabulary words, and when Nomon was used in a picture selection task. Overall, our results show that Nomon may provide a more efficient, and more flexible, method for rate-limited users to control their computer via a single switch.

\begin{acks}
We thank the two switch users, who provided important feedback and evaluated the Nomon interface. We are grateful to Bill Donegan and Mick Donegan for organizing several sessions with the staff of SpecialEffect, who provided invaluable feedback towards the final design of Nomon. We thank Heidi Koester for sharing data and useful comments. We thank the Ace Centre – including Will Wade – for helpful conversations. This work was supported in part by the Seth Teller Memorial Fund to Advance Technology for People with Disabilities, a Peter J. Eloranta Summer Undergraduate Research Fellowship, the MIT Intelligence Quest, and the NSF under Grant No. IIS-1750193.
\end{acks}

	\bibliographystyle{ACM-Reference-Format}
	\bibliography{references}

	\newpage
	\appendix
	


\onecolumn

{\Huge Supplemental Material: A Performance Evaluation of Nomon: A Flexible Interface for Noisy Single-Switch Users}


\section{Simulations of RCS for Optimization}
\label{section:Simulations of Usage of Software} 
Row Column Scanning (RCS) keyboards have various parameters that control the presentation of characters and word completions on its interface. As discussed in \cite{koesterscanwizzard}, these parameters directly influence performance in terms of text entry metrics (e.g., text entry rate and click load) . We developed a simulation framework for RCS keyboards to optimize these parameters to provide the highest text entry rate. 

\subsection{RCS Model for a Simulated User} 
 Our simulated user is similar to Mankowski et al.~\cite{koesterrcsmodel}. We added small modifications to account for click-timing noise so that we can directly compare to Nomon. We have validated the final optimized parameters against the state of the art as described in Section 3.1.3 of the main text.
 In order to output a switch activation time, the simulated user (SU) needs a target word or character to select. The SU types each word in the target phrase one character at a time. If the word currently being typed is in the word predictions, this word prediction is the target. Otherwise, the next character in the current word is the target. After it has a target, the SU needs to determine when to send a switch input to select this target based on the current state of the keyboard.

First, the SU calculates the time until the optimal switch input, $t^*$, for an ideal user.
In RCS, an ideal user would press midway through the highlight time of the desired row or column. The SU calculates the time until $t^*$ by multiplying the scan delay by number of scans between the currently highlighted row/column and the target row/column. Given a target row $r_{\mathrm{T}}$ and column $c_{\mathrm{T}}$, a currently highlighted row~$r$  and column $c$, and a scan delay $D$ (measured in seconds), $t^*$ can be calculated from the following equations: %
\[
t^* = \begin{cases}
(r_{\mathrm{T}} - r) \cdot D & \text{if row scan and } r_{\mathrm{T}} \geq r\\
(R - r + r_{\mathrm{T}}) \cdot D & \text{if row scan and } r_{\mathrm{T}} < r\\
(c_{\mathrm{T}}-c) \cdot D & \text{if column scan and } c_{\mathrm{T}} \geq c\\ 
(C - c + c_{\mathrm{T}}) \cdot D & \text{if column scan and } c_{\mathrm{T}} < c\\
\end{cases}
,
\]

where $R$ is the number of rows, $C$ is the number of columns, $r, r_{\mathrm{T}} \in \{1, \dots, R\}$, and $c, c_{\mathrm{T}} \in \{1, \dots, C\}$. 

For example, if the simulated user has to select the third row, and there a five rows in total, $r_{\mathrm{T}} = 3$, $R=5$, and  
\[
t^* = \begin{cases}
2D & \text{if } r = 1\\
D & \text{if } r = 2\\
0 & \text{if } r = 3\\
4D & \text{if } r = 4\\
3D & \text{if } r = 5\\
\end{cases}.
\]     

\begin{figure}[h]
	\centering
	\includegraphics[width=95mm]{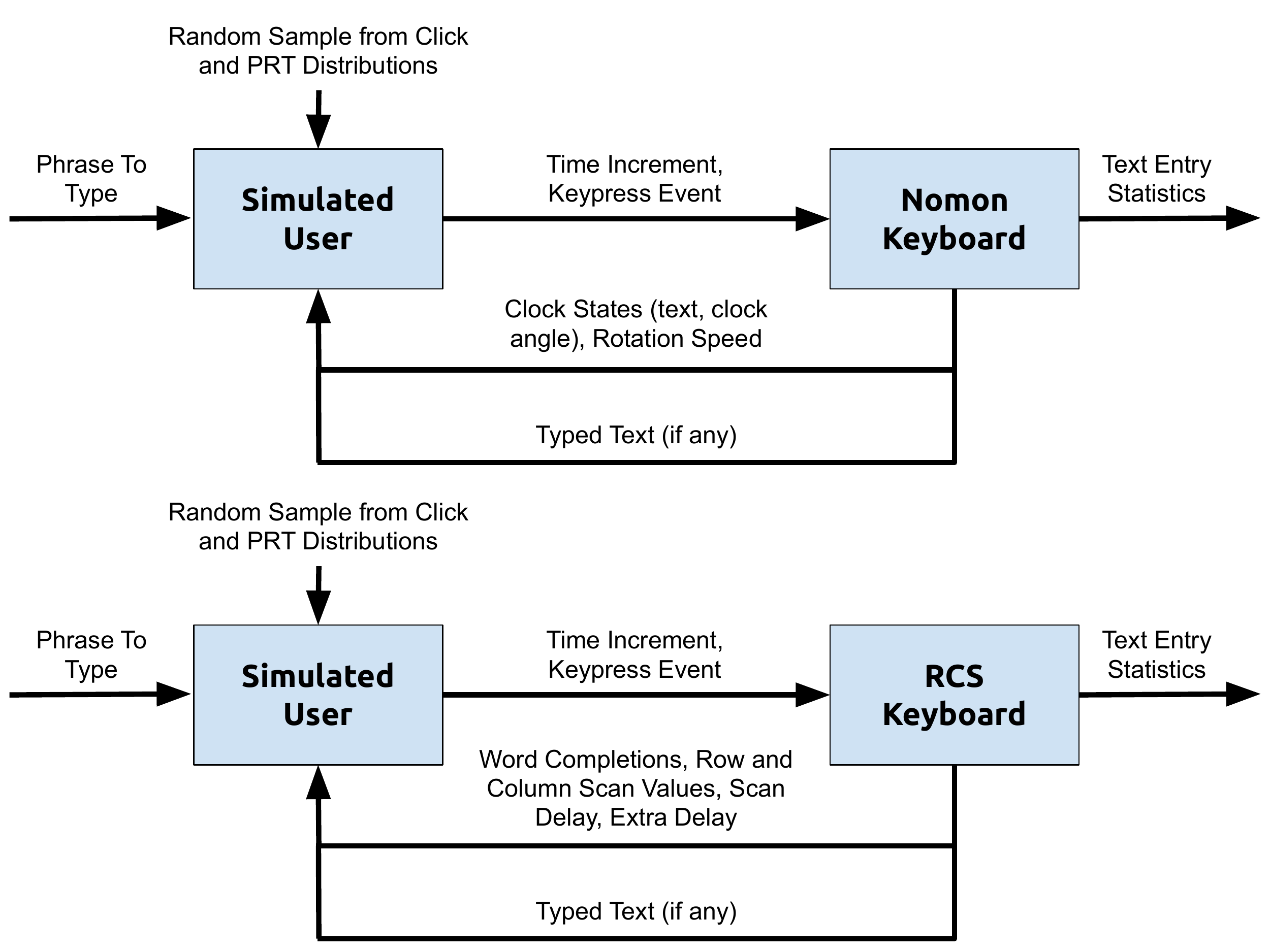}
	\caption{Diagram showing the operation of the Nomon simulated user (top) and the RCS simulated user (bottom).}
	\label{figure:Simulated User Diagram}
\end{figure}

However, an actual user will rarely press exactly at $t^*$ and our SU needs a mechanism to model user input noise. We model an imperfect user by adding to $t^*$ an $iid$ sample $t_c$ from the click distribution of an expert user. The click distribution of a user is an empirical likelihood of their click times relative to the optimal click time. Click distributions for RCS users are similar to Nomon (discussed in the main paper, Section 3), but they are measured relative to the midway time of a row or column scan instead of noon.

After $t^*$ and $t_c$ are calculated, the SU increments the keyboard application by $t^*+t_c$ seconds and sends a switch input. If $t_c$ is too positive or negative, the SU will select an incorrect row or column, making an error.  In such cases we assume that the user will attempt to undo their errors. In the case of an an erroneous row selection, the user can wait for 2R scans to undo their selection, in which case 2RD will also have to be added to $t^*$. In the case of an erroneous column selection, the user can select the undo cell. The user is allowed a finite number of attempts to correct an error (we allowed two attempts in our simulation). If the SU is unable to correct their error, the simulation will result in a failure and move on to the next word. This failure leaves the incorrect selection in place, resulting in a non-zero final error rate for the phrase. 

\subsection{Parameter Search} 
To determine the optimal configuration the RCS, we ran simulations over a wide range of parameter values. We considered the following parameters:
\begin{itemize}
    \item Word Predictions Max Count ($W_{\text{max}}$) --- This parameter is the total number of word predictions allowed on the screen at a time. We considered values $W_{\text{max}} \in \{1, 2, ... ,18\}$.

    \item Word Prediction Location --- This parameter can take values of \textit{top} or \textit{bottom} and controls whether the word predictions are above or below the characters in the grid.

    \item Key Sorting --- This parameter can take values of \textit{alphabetical} or \textit{frequency} and determines the order of characters in the grid. The alphabetical layout sorts characters alphabetically across the rows and columns. The frequency based layout places more common letters in English closer to the top-left of the grid to reduce the scan time to them.
\end{itemize}

\subsection{Results} 
\begin{figure}[tb]
	\centering
	\includegraphics[width=90mm]{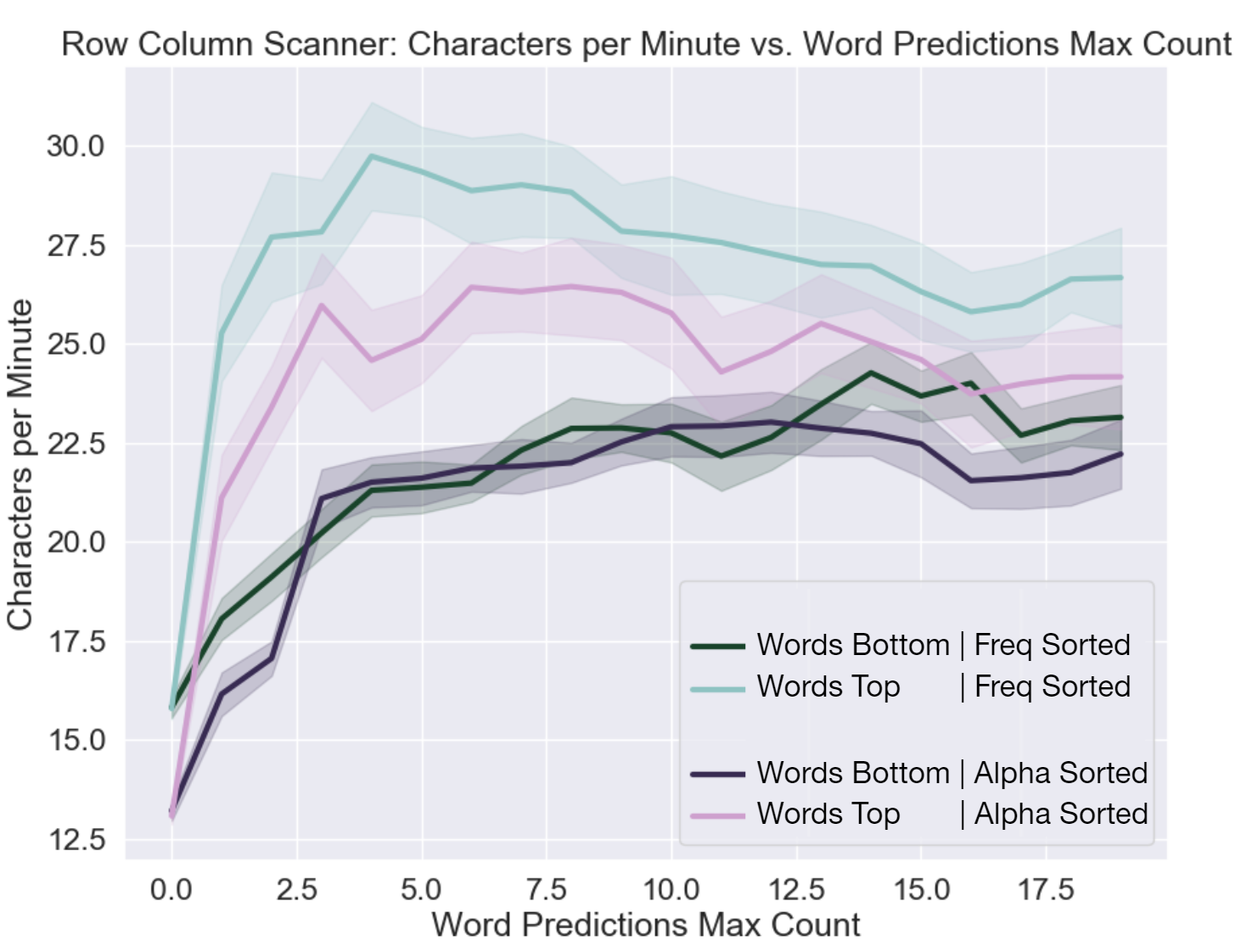}
	\caption{Simulated text entry rate in RCS for breath of $W_{max}$, Word Prediction Location, and Key Sorting parameters.}
	\label{figure:Row Col Entry Rate Sim}
\end{figure}

We seek to optimize the parameters $W_{max}$, Word Prediction Location, and Key Sorting to provide the highest text entry rate. Looking at Figure \ref{figure:Row Col Entry Rate Sim}, the combination of placing word predictions at the top and using frequency sorting provides the highest text entry rate for all values of $W_{max}$. The maximum text entry rate is achieved at values of $W_{max} \approx 5$, as adding too many word predictions hurts RCS. We conclude that the optimal layout for the RCS keyboard is with word predictions at the top, characters sorted by frequency, and $W_{max} = 7$ (since this value fills an entire row with word predictions, and there is significant overlap between the error bars at $W_{max} = 5$ and 7).

\section{Nomon Applications Beyond Text Entry}
\subsection{Drawing}
The Nomon drawing application was created as a proof of concept in \cite{nomonthesis}. This application highlights Nomon's ability to handle large numbers of options (on the order of 400 in this example) without a prohibitive increase in selection time. An experienced Nomon user was able to select among the clocks in this drawing task with an average of 3 clicks \cite{nomonthesis}. Under an uninformed prior (as in this task), previous work has shown that the number of switch clicks required to select a target in Nomon scales logarithmically with the number of options \cite{nomonthesis}. Though our picture selection task in Section 4.4 of the main text has significantly fewer clocks (about 60), we found participant's click load remained low at around 3.5 clicks per selection even using the noisy webcam switch. Future work could explore how this trend holds with many more clocks as in the Nomon drawing application.

\begin{figure}[tb]
    \centering
    \includegraphics[width=0.8\linewidth]{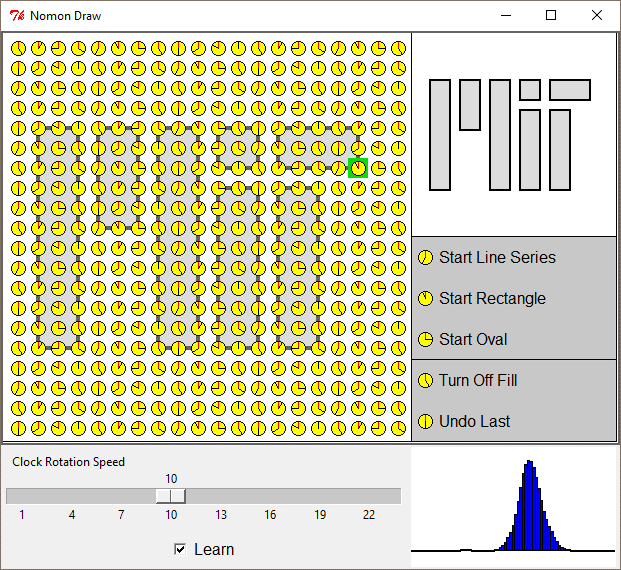}
    \caption{A screenshot of the Nomon drawing application. The application allows users to draw lines, rectangles, and ovals over a grid of points. For example, to draw a rectangle, the user first selects the "Draw Rectangle" clock in the right sidebar. They then select one of the clocks among the grid of clocks to define the starting corner of the rectangle. Finally, they select another clock among the grid to define the rectangle's ending corner.}
    \label{fig:nomon draw}
\end{figure}
\subsection{Web Browsing and General Operating System Control}
\begin{figure}[tb]
    \centering
    \includegraphics[width=120mm]{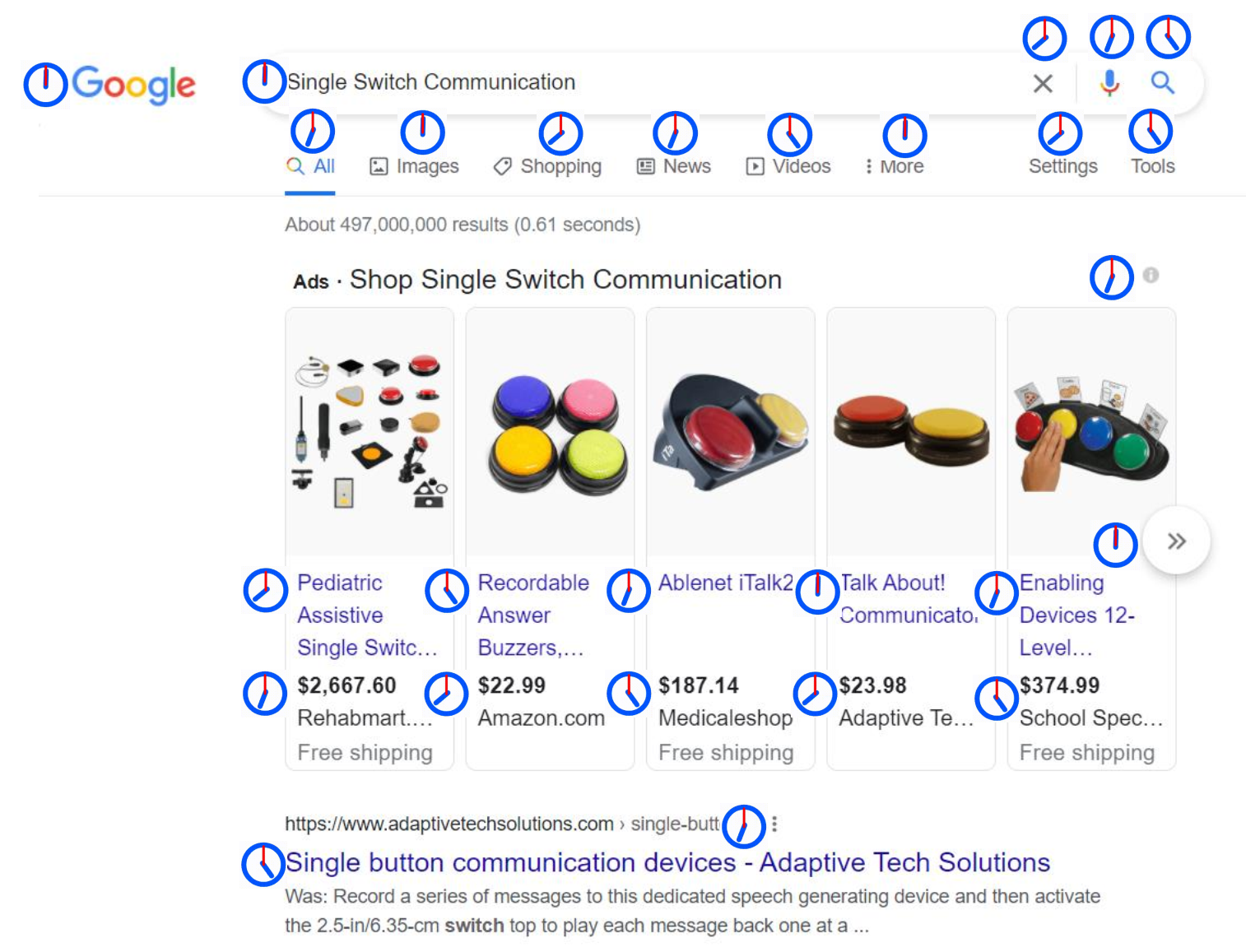}
    \caption{A mock-up of Nomon applied to a web browsing task. A Nomon clock could be positioned next to every item on a webpage (buttons, hyperlinks, checkboxes, textboxes, etc) to make the webpage fully accessible with a single switch. }
    \label{fig:nomon browser}
\end{figure}

Nomon does not restrict options to a grid like many scanning-based interfaces. This flexibility could allow Nomon to be integrated seamlessly with GUI applications like web browsers and operating system control. Figure \ref{fig:nomon browser} shows a mock-up of how a Google search result page could be outfitted with Nomon clocks to allow single-switch interaction with the webpage. The results from our picture selection task in Section 4.4 of the main text show that Nomon can handle the large number of clocks that may be required for GUI control without dramatically increasing selection time. One can imagine that the number of clocks required for GUI control can add up quickly since each tab, hyperlink, button, menu option, textbox, etc. must be accessible.

\section{Participant Free Responses Final Session}

\paragraph{Free Response from Nomon Trial with Switch User:}

``My first impressions of Nomon (prior to this study) were that it was a silly work around to avoid row-column scanning. As I use Nomon for text entry, I notice a sense of direct selection (though technically it is not) akin to eye gaze interfaces. One important difference is that I did NOT experience the same eye strain/fatigue often associated with eye gaze mouse pointer navigation.

``After copying the second sentence, I observed more word predictions showing up as choices. This is where I see some real potential for increased typing rate (in terms of words per minute). Nomon is distinctly different from traditional scanning and may offer an easier path to higher text entry rates. My typing rate with EZ Keys row-column scanning is 13 wpm with the an insane step-time of 100ms. Making intentional menu selections at that speed requires a hair-trigger switch and internalized knowledge of the menu patterns (not every user wants typing to be an extreme sport, eh? ).

``I’m looking forward to pushing the envelope ....''


\begin{table}[h]
	\begin{tabular}{ll}
		\toprule
		\multicolumn{2}{c}{\textbf{Nomon Free Responses Session 9}} \\
		\hline
		\multicolumn{1}{c|}{Comment} & \multicolumn{1}{c}{Themes} \\
		\hline
		\multicolumn{1}{p{10cm}|}{I like that this interface has lots of room for suggestions so typing feels faster, and there's less down time waiting for the clock to come around for each letter. 
		}                                                                                                                 & \multicolumn{1}{p{3.2cm}}{downtime (+), \newline word completions (+)} \\
		\hline
		\multicolumn{1}{p{10cm}|}{I like how the predictive words are organized by the next letter you might be typing.
		} 
		& \multicolumn{1}{p{3.2cm}}{word completions (+)} \\
		\hline
		\multicolumn{1}{p{10cm}|}{I think I would like more contrast in the alphabet. Maybe a bigger font.
		}
		& \multicolumn{1}{p{3.2cm}}{layout (-)} \\
		\hline
		\multicolumn{1}{p{10cm}|}{I think I figured out that there is some kind of rhythm if you pick the right clock that lets you rapidly pick that letter or word. It seems that the faster clocks are actually easier to get the hang of.
		}                                                                                                                 & \multicolumn{1}{p{3.2cm}}{rhythm (+)} \\
		\hline
		\multicolumn{1}{p{10cm}|}{What I like about this keyboard, is that it gives more full word options since you can pick any of the letters. There's here's less down time (waiting for the right row to highlight).
		} 
		& \multicolumn{1}{p{3.2cm}}{downtime (+), \newline word completions (+)} \\
		\hline
		\multicolumn{1}{p{10cm}|}{This keyboard almost has a rhythm to its usage; you're never waiting for more than one clock rotation before you make a selection, but you have the ability to wait a rotation without it feeling like you've messed up. There's simultaneously not enough waiting time for me to lose too much focus on an individual task.
		}
		&\multicolumn{1}{p{3.2cm}}{downtime (+), \newline rhythm (+)} \\
		\hline
		\multicolumn{1}{p{10cm}|}{It definitely seems a lot faster to write with this keyboard. I like the fact that you aren't wasting your time when typing less frequently used letter. It sometimes can get a little tiring making constant clicks.,
		}
		& \multicolumn{1}{p{3.2cm}}{downtime (+)} \\
		\hline
		\multicolumn{1}{p{10cm}|}{I liked predictable locations of letters and words, if I missed a chance to activate next time around came quickly not too much waiting. I disliked not being able to see the difference between blue and black easily and if it turned black I didnt know the fated way to correct that.
		}
		& \multicolumn{1}{p{3.2cm}}{downtime (+)} \\
		\hline
		\multicolumn{1}{p{10cm}|}{I like how easy it is to select each letter, due to the selection of one not being dependent on something else, like selecting a row.
		}        
		& \multicolumn{1}{p{3.2cm}}{downtime (+), \newline flexibility (+) } \\
		\hline
		\multicolumn{1}{p{10cm}|}{I don't like how the clocks start over at a random spot every time you press it because it catches you off guard and if it starts really close to the top then you have to wait for it to come back around the next time
		}
		& \multicolumn{1}{p{3.2cm}}{mechanics (-) } \\
		\hline
		\multicolumn{1}{p{10cm}|}{Unlike the row/col interface, all letters are always accessible. This means that phrases with less common letters (e.g. 'b' or 'z') can be easier and quicker to type (especially when the word of interest is irregular or a typo and doesn't show up as a suggested word). However, this comes at a cost since the interface is more cluttered and there is a lot of information on the screen that users need to pay attention to.
		}                                                                                     
		& \multicolumn{1}{p{3.2cm}}{flexibility (+), \newline layout (-)} \\
		\hline
		\multicolumn{1}{p{10cm}|}{I liked how many word predictions there were. It seems like there are more on this interface.
		}
		& \multicolumn{1}{p{3.2cm}}{word completions (+) } \\
		
		\bottomrule               
	\end{tabular}
\end{table}

\begin{table}[h]
	\begin{tabular}{ll}
		\toprule
		\multicolumn{2}{c}{\textbf{RCS Free Responses Session 9}} \\
		\hline
		\multicolumn{1}{c|}{Comment} & \multicolumn{1}{c}{Themes} \\
		\hline
		\multicolumn{1}{p{10cm}|}{I disliked that one wrong move makes the scanner lock onto a row or select a letter, where the other one required more precision on the temporal aspect. This creates a huge frustration when typing.
		}                                                                                                                 & \multicolumn{1}{p{3.2cm}}{mechanics (-), \newline frustration (-)} \\
		\hline
		\multicolumn{1}{p{10cm}|}{At the speed I'm at when there's no options in the predictive row I don't have time to select the first row of letters before the selection bar scrolls down.
		} 
		& \multicolumn{1}{p{3.2cm}}{mechanics (-)} \\
		\hline
		\multicolumn{1}{p{10cm}|}{Word prediction was good.
		}
		& \multicolumn{1}{p{3.2cm}}{word completions (-)} \\
		\hline
		\multicolumn{1}{p{10cm}|}{Once you get an idea of timings and such, it's very easy to plan out how to type things and get faster at it.
		}                                                                                                 & \multicolumn{1}{p{3.2cm}}{mechanics (+)} \\
		\hline
		\multicolumn{1}{p{10cm}|}{Once you learn the count times between the letters it makes it easier to not have as many mistakes
		} 
		& \multicolumn{1}{p{3.2cm}}{mechanics (+)} \\
		\hline
		\multicolumn{1}{p{10cm}|}{This keyboard feels like more of a chore than Keyboard A.
		}
		&\multicolumn{1}{p{3.2cm}}{mechanics (-)} \\
		\hline
		\multicolumn{1}{p{10cm}|}{It feels really slow to type with this keyboard and this slowness can amount to frustration. It's definitely easier on the physical and mental stress since there's more time in between each head tilts.
		}
		& \multicolumn{1}{p{3.2cm}}{speed (-), \newline frustration (-)} \\
		\hline
		\multicolumn{1}{p{10cm}|}{did not like that the second line comes and goes so fast I always missed selecting it.
		}
		& \multicolumn{1}{p{3.2cm}}{mechanics (-)} \\
		\hline
		\multicolumn{1}{p{10cm}|}{This way of typing is a lot more straight-forward than the clocks, and there is a lot less to focus on. This makes it much easier (less cognitive load), but I think that it can be a little slow. As I use this interface more and more, I feel like I need to keep increasing the speed.
		}        
		& \multicolumn{1}{p{3.2cm}}{speed (-), \newline mechanics (+) } \\
		\hline
		\multicolumn{1}{p{10cm}|}{I find it frustrating how long it takes wait for a row to scan through twice if accidentally hit. Even with the scan speed up it's annoying.
		}        
		& \multicolumn{1}{p{3.2cm}}{downtime (-), \newline frustration (+) } \\
		\hline
		\multicolumn{1}{p{10cm}|}{I like that you can change the speed after each sentence, it is not as tedious as the clock keyboard
		}
		& \multicolumn{1}{p{3.2cm}}{mechanics (+) } \\
		\hline
		\multicolumn{1}{p{10cm}|}{I liked that you could adjust the speeds of the scanning. 
		}
		& \multicolumn{1}{p{3.2cm}}{mechanics (+) } \\
		
		\bottomrule               
	\end{tabular}
\end{table}

\begin{table}[h]
	\begin{tabular}{l|l|l}
		\toprule
		\multicolumn{2}{c}{\textbf{Preferred Software Responses Session 9}} \\
		\hline
		\multicolumn{1}{p{1.2cm}|}{Preferred} & \multicolumn{1}{c|}{Comment} & \multicolumn{1}{c}{Themes} \\
		\hline
		Nomon &
		\multicolumn{1}{p{10cm}|}{I liked that Keyboard A is more forgiving with errors, faster to type, and less downtime waiting for scanning.
		}                                                                        
		& \multicolumn{1}{p{1.6cm}}{error corr, \newline speed, \newline downtime } \\
		\hline
		Nomon &
		\multicolumn{1}{p{10cm}|}{I feel like I have more agency and can type faster with keyboard a.
		} 
		& \multicolumn{1}{p{1.6cm}}{speed, \newline flexibility} \\
		\hline
		Nomon &
		\multicolumn{1}{p{10cm}|}{I think, over time, I would be much faster on A. 
		}
		& \multicolumn{1}{p{1.6cm}}{speed} \\
		\hline
		Nomon &
		\multicolumn{1}{p{10cm}|}{I think it gives more flexibility as you can select any letter at any given time, without waiting for the right row to highlight. I also like how for each letter it gives a selection of words so it's easier to find the right word.
		}                                                                        
		& \multicolumn{1}{p{1.6cm}}{flexibility, \newline layout} \\
		\hline
		Nomon &
		\multicolumn{1}{p{10cm}|}{I think the clocks allow the user to correct their mistakes more quickly, which cuts down on a lot of waiting time. The grid you need to learn a bit more like the layout and how to count between rows and columns, but I think they are equally good choices.
		} 
		& \multicolumn{1}{p{1.6cm}}{flexibility, \newline downtime} \\
		\hline
		Nomon &
		\multicolumn{1}{p{10cm}|}{Keyboard A is more forgiving due to the ability to select things quickly or choose not to select without the seeming penalty of Keyboard B's long cycle time. It's alphabetical layout is also more intuitive; alphabetical wouldn't necessarily work for Keyboard B functionally, but the combination of that layout and cycling method make Keyboard A my preference overall.
		}
		&\multicolumn{1}{p{1.6cm}}{flexibility, \newline layout, \newline downtime} \\
		\hline
		Nomon &
		\multicolumn{1}{p{10cm}|}{It feels like you can type a lot quicker and doesn't seem to drag on as Keyboard B does.
		}
		& \multicolumn{1}{p{1.6cm}}{speed, \newline downtime} \\
		\hline
		Nomon &
		\multicolumn{1}{p{10cm}|}{Location of things and timing is predictable. seems faster and easier to correct than B. Spend lots of time with B waiting not much idle time with A.
		}
		& \multicolumn{1}{p{1.6cm}}{error corr, \newline speed, \newline downtime, \newline layout} \\
		\hline
		Nomon &
		\multicolumn{1}{p{10cm}|}{The clocks felt more user friendly, by being both more user friendly and faster to navigate. 
		}        
		& \multicolumn{1}{p{1.6cm}}{speed, \newline mechanics} \\
		\hline
		Nomon &
		\multicolumn{1}{p{10cm}|}{I feel as if this method is more accurate and it is faster than the other. it doesn't take as long for the clock to spin around so you can press the key that you would like. more efficient
		}        
		& \multicolumn{1}{p{1.6cm}}{speed, \newline downtime} \\
		\hline
		RCS &
		\multicolumn{1}{p{10cm}|}{Initially, I strongly preferred Keyboard B since it was immediately easy to use. However, as I had more time to practice both interfaces, I became a lot more comfortable with both. I still think I slightly prefer Keyboard B because it is easier for me (a little less cognitive load), but that is only true when I am able to increase the speed and there are not a lot of [OOV] words in the phrases. For me, Keyboard B shines when the words in the phrase are easily predictable and the speed is higher, otherwise it can seem cumbersome and slow.
		}
		& \multicolumn{1}{p{1.6cm}}{mechanics} \\
		\hline
		Nomon &
		\multicolumn{1}{p{10cm}|}{Even though I prefer B, I can get to full words faster on A.  Also, as I got better I could speed the clocks up and go even faster.
		}
		& \multicolumn{1}{p{1.6cm}}{speed} \\
		\hline
		Nomon &
		\multicolumn{1}{p{10cm}|}{There are a lot of word predictions with the clocks, and I feel like it8s faster to type this this interface.
		}
		& \multicolumn{1}{p{1.6cm}}{word pred, \newline speed} \\
		\bottomrule               
	\end{tabular}
\end{table}




\end{document}